\documentclass[english]{iopart}
\pdfoutput=1
 \usepackage{graphicx}

\newcommand{\beq}{\begin{equation}}
\newcommand{\eeq}{\end{equation}}

\newcommand{\beqr}{\begin{eqnarray}}
\newcommand{\eeqr}{\end{eqnarray}}
\newcommand{\barr}{\begin{array}}
\newcommand{\earr}{\end{array}}
\newcommand{\bal}{\begin{align}}
\newcommand{\eal}{\end{align}} 
\newcommand{\bmu}{\begin{multline}}
\newcommand{\emu}{\end{multline}}

\begin{document}

\centerline{\bf \Large Integrability vs non-integrability}
\centerline{\bf \Large Hard hexagons and hard squares compared}

\author{  M. Assis$^1$, J.L. Jacobsen$^{2,3}$, I. Jensen$^4$, J-M. Maillard$^5$ and B.M. McCoy$^1$ }
\address{$^1$ CN Yang Institute for Theoretical Physics, State
  University of New York, Stony Brook, NY, 11794, USA} 
\address{$^2$ Laboratoire de Physique Th{\'e}orique, {\'E}cole Normale
  Sup{\'e}rieure, 24 rue Lhomond, 75231 Paris Cedex, France}
\address{$3$ Universit{\'e} Pierre et Marie Curie, 4 Place Jussieu,
  75252 Paris, France}
\address{$^4$ ARC Center of Excellence for Mathematics and Statistics
of Complex Systems, Department of Mathematics and Statistics, The
University of Melbourne, VIC 3010, Australia}
\address{$^5$ LPTMC, UMR 7600 CNRS, 
Universit\'e de Paris, Tour 23,
 5\`eme \'etage, case 121, 
 4 Place Jussieu, 75252 Paris Cedex 05, France} 

\begin{abstract}

In this paper we compare the integrable hard hexagon model with the
non-integrable hard squares model by means of partition function roots
and transfer matrix eigenvalues. We consider partition functions for
toroidal, cylindrical, and free-free boundary conditions up to
sizes $40\times40$ and transfer matrices up to 30 sites. For all boundary
conditions the hard squares roots are seen to
lie in a bounded area of the complex fugacity plane along with the
universal hard core line
segment on the negative real fugacity axis. The
density of roots on this line segment matches the derivative of the
phase difference between the eigenvalues of largest (and equal) moduli and 
exhibits much greater structure than the corresponding density 
of hard hexagons. We also study the
special point $z=-1$ of hard squares  where all eigenvalues have unit
modulus, and we give several conjectures for the value at $z=-1$ of
the partition functions.
 
\end{abstract}

\noindent {\bf AMS Classification scheme numbers}: 34M55, 
47E05, 81Qxx, 32G34, 34Lxx, 34Mxx, 14Kxx 
\vskip .5cm


\vspace{.2in}

\section{Introduction}

There is a fundamental paradox in the practice of theoretical physics.
We do exact computations on integrable systems which have very special
properties and then apply the intuition gained to  generic systems
which have none of the special properties which allowed the exact
computations to be carried out. The ability to do exact computations
relies on the existence of sufficient symmetries which allow the system
to be solved by algebraic methods. Generic systems do not possess such
an algebra and the distinction between integrable and non-integrable
may be thought of as the distinction of algebra versus analysis.

This paradox is vividly illustrated by the two dimensional Ising
model. In zero magnetic field Onsager 
\cite{ons} computed the free energy by means of exploiting the
algebra which now bears his name.
 On the other hand in 1999 Nickel \cite{nic1,nic2} analyzed the 
expansion of the
susceptibility at zero magnetic field for the isotropic Ising model on
the square lattice and discovered that as a function
of the variable $s=\sinh 2E/k_BT$ the susceptibility has a dense set of
singularities on the circle $|s|=1$ which is the same location as 
the thermodynamic limit of the locus of zeros of the finite lattice 
partition function. From
this Nickel concluded that the curve of zeros is a natural boundary of
the susceptibility in the complex $s$ plane. This is a  phenomenon of
analysis  not seen in any previously solved statistical system. 
Further study of
this new phenomenon has been made by Orrick, Nickel, Guttmann and 
Perk \cite{ongp} and in \cite{cgnp} the phenomenon
of the natural boundary was studied on the triangular lattice. 
However the implication of these results for other models has not 
been investigated.

The hard square and hard hexagon models can be obtained from the Ising
model in a magnetic field $H$ in the limit $H\rightarrow \infty$
for the square and triangular lattices respectively, and thus it is
natural to  study the question of analyticity in these two models.
However, unlike the Ising model at $H=0$ where both the square and
triangular lattices have been exactly solved, the hard hexagon model
is exactly solved \cite{baxterhh}-\cite{baxpaul} whereas the hard square model is not. Thus, the
comparison of these two models is the ideal place to study the relation
of integrability to the analyticity properties of the free energy in
the complex plane.

Three different methods may be used to study the non-integrable hard
square model: Series expansions of the free energy in the
thermodynamic limit, transfer matrix eigenvalues for chains of finite
size $L_h$ and zeros of partition functions on the $L_v\times L_h$
lattices of finite size and arbitrary aspect ratio $L_v/L_h$.

Series expansions of  
the partition function per site $\kappa(z)$ of  the hard square model
\cite{gf}-\cite{kam} of up to 
92 terms \cite{chan} and analysis of transfer matrix eigenvalues \cite{ree}
for chains of up to 34 sites \cite{gb} show that $\kappa(z)$ has a
singularity on the
positive $z$-axis \cite{gb}
\begin{equation}
z_c=3.79625517391234(4)
\end{equation}
and a singularity on the negative $z$-axis  \cite{tony,jensen}
\begin{equation}
z_d=-0.119338886(5)
\end{equation}
The hard hexagon model has two singular points at
\cite{baxterhh}-\cite{baxpaul}
\begin{eqnarray}
&&z_{c;hh}=\frac{11+5{\sqrt 5}}{2}= 11.09016\cdots\nonumber\\
&&z_{d;hh}=\frac{11-5{\sqrt 5}}{2}=-0.09016\cdots
\end{eqnarray}
For hard squares, series expansions \cite{gf}-\cite{kam}  have been used to
estimate the leading critical exponents at $z_c$ and $z_d$, and 
correction to scaling exponents have been estimated as well. For hard
hexagons there are no singular points of the free energy  other than 
$z_{c;hh},~ z_{d;hh},~\infty$.  It is not known if there are any 
further singular points for hard squares. In \cite{gb} the singularity 
at $z_c$ is determined to be in the Ising
universality class and in \cite{jensen} the first two exponents at
$z_d$ are shown to agree with those of the Lee-Yang edge and hard hexagons.
However these long series expansions have not given information about
additional higher order singularities at $z_c$ and $z_d$ or 
singularities which may occur at other values of $z$.

In 2005 a very remarkable property of hard squares, which is not shared
by hard hexagons, was discovered \cite{fse} by means of studying 
the eigenvalues
of the transfer matrix for finite size systems
\cite{fse}-\cite{baxneg}. These studies discovered that at the 
value of the fugacity $z=-1$ all eigenvalues
of the transfer matrix with cylindrical boundary conditions have unit 
modulus and the partition function of the $L_h\times L_v$ lattice 
with toroidal boundary conditions depends on
divisibility properties of $L_v$ and $L_h$. However, the free energy
for these boundary conditions in the thermodynamic limit is zero.
For the lattice oriented at $45^o$, on the other hand, for cylindrical
boundary conditions  of the transfer matrix, there are some eigenvalues
which do not have unit modulus \cite{jon2} and for free boundary
conditions of the transfer matrix with $L_h\equiv 1~({\rm mod}~ 3)$ 
all roots of the characteristic equation are zero and thus the 
partition function vanishes.

In \cite{assis} we computed for hard hexagons   
the  zeros of the  partition function for $L\times L$ lattices 
with cylindrical and toroidal boundary conditions as large as
$39\times 39$ and the eigenvalues
of the transfer matrix with cylindrical boundary conditions. 
For these cylindrical transfer matrices both momentum and parity are conserved,
and for physical (positive) values of $z$ the maximum eigenvector is
in the sector of zero momentum positive parity $P=0^+$.
From these cylindrical transfer matrices we computed 
the equimodular curves where there are two eigenvalues of the row transfer
matrix of (equal) maximum modulus both in the sector $P=0^+$ and for
the full transfer matrix.

In this paper we extend our study of
partition function zeros and transfer matrix equimodular curves to hard
squares for systems as large as $40\times 40$ and compare them with 
corresponding results for hard hexagons \cite{assis}. There are
many differences between these two systems which we analyze in detail.
In addition to the transfer matrix with cylindrical boundary
conditions we also introduce the transfer matrix with free boundary
conditions. Thus we are able to give two different transfer matrix
descriptions for the partition function zeros of the cylindrical
lattice. For hard hexagons there is strong evidence that this boundary
condition preserves integrability.

In section \ref{recall} we recall the relation between finite size
computations in the complex plane of zeros of $L\times L$ lattices 
and eigenvalues of the $L$ site transfer matrix. 
In section \ref{compare} we make a global comparison in the complex $z$
plane of the equimodular curves and partition function zeros of hard
squares with hard hexagons. In section \ref{negz} we make a more
refined comparison on the negative $z$ axis.

The comparisons presented in sections \ref{compare} and \ref{negz} 
reveal many 
significant differences between hard squares and hard hexagons which
we discuss in detail in section \ref{discussion}. We conclude in
section \ref{conclusions} with a presentation of 
potential analyticity properties of hard squares which can be different
from hard hexagons. 

In \ref{appa}  we tabulate the factored characteristic polynomials of
the transfer matrix at the point $z=-1$ and  the multiplicity of
the eigenvalue $+1$. We also give formulas for the growth of the orders
of the transfer matrices, where such a formula is known, and for all 
cases the asymptotic growth is given by $N_G^{L_h}$ where $N_G$ is the
golden ratio. 

In \ref{appb} we consider the partition function values at $z=-1$ 
on $L_v\times L_h$ lattices  for the torus, cylinder, free-free rectangle,
M{\"o}bius band and Klein bottle boundary conditions. We give
generating functions for the sequences of values of the partition
function of the $L_v\times L_h$ lattice as a function of $L_v$ 
and find that almost 
all sequences of values are repeating. We conjecture that 
along the periodic $L_v$ direction 
(including twists for the M{\"o}bius band and
Klein bottle cases) the sequences will always be
repeating. Furthermore, for the torus and the cylinder (along the
periodic $L_v$ direction), we conjecture that the generating functions are
given by the negative of the logarithmic derivative of the
characteristic polynomial of their transfer matrices at
$z=-1$. This allows us to conjecture the periods of their repeating
sequences. Finally, for
the M{\"o}bius band (along the periodic $L_v$ direction)  
and Klein bottle we conjecture that their
generating functions are  the logarithmic derivative of 
products of factors $(1-x^{n_i})^{m_j}$, where $n_i,~m_j$ are integers.

\section{Formulation}
\label{recall}

The hard square lattice gas is defined by a (occupation) variable 
$\sigma=0,1$ at each site of a square lattice with the restriction 
that no two adjacent sites can have the values $\sigma=1$ (i.e. the
gas has nearest neighbor exclusion).
The grand partition function on the finite $L_v\times
L_h$ lattice is defined as the polynomial
\begin{equation}
Z_{L_v,L_h}(z)=\sum_{n=0}z^ng(n;L_v,L_h).
\end{equation}
where $g(n;L_v, L_h)$ is the number of hard square 
configurations which have $n$ occupied sites.
These polynomials can be characterized by their zeros $z_j$ as
\begin{equation}
Z_{L_v,L_h}(z)=\prod_j(1-z/z_j),
\end{equation}
where  $z_j$ and the degree of the polynomial will depend on the boundary condition imposed on the
lattice. This formulation of the partition function as a polynomial 
is completely general for lattice models with arbitrary interactions.    

\begin{figure}[h!]
\begin{center}
\hspace{0cm} \mbox{
\begin{picture}(100,100)
\put(20,0){\includegraphics[width=3cm]{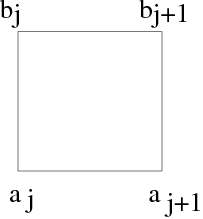}} 
\end{picture}
}
\end{center}
\caption{Boltzmann weights for the transfer matrix of hard squares}
\label{fig:bw}
\end{figure}

The partition function for hard squares may also be expressed in
terms of the transfer matrix formalism. 
For the cylindrical transfer matrix with  periodic boundary conditions 
in the horizontal direction, the transfer matrix for hard squares is defined as
\begin{eqnarray}
 \hspace{-0.9in}&& \qquad \quad  \quad 
T_{C\{b_1,\cdots b_{L_h}\},\{a_1,\cdots.a_{L_h}\}}(z;L_h)
\,\,  =\,\,\, \,   \prod_{j=1}^{L_h}\, W(a_j,\, a_{j+1};\, b_j,\, b_{j+1}),  
\label{tmat}
\end{eqnarray}
where the local Boltzmann weights
$W(a_j,a_{j+1};b_j,b_{j+1})$ for hard squares of figure \ref{fig:bw}
 may be written as
\begin{equation}
\hspace{-0.3in} W(a_j,a_{j+1};b_j,b_{j+1})=0 ~~{\rm for}~~
a_ja_{j+1}=a_{j+1}b_{j+1}=b_jb_{j+1}=a_jb_j=1
\label{bw1}
\end{equation}
with $a_{L_h+1}\equiv a_1,~b_{L_h+1}\equiv b_1$ and otherwise
\begin{equation}
W(a_j,a_{j+1};b_j,b_{j+1})=z^{b_j}.
\label{bw2}
\end{equation}

For the transfer matrix with free boundary conditions 
\begin{eqnarray}
 \hspace{-0.9in}&& 
T_{F\{b_1,\cdots b_{L_h}\},\{a_1,\cdots.a_{L_h}\}}(z;L_h)=\nonumber\\
&&  \left(\prod_{j=1}^{L_h-2} W(a_j,\, a_{j+1};\, b_j,\, b_{j+1})\right)  
W_F(a_{L_h-1},a_{L_h};b_{L_h-1},b_{L_h}),
\label{tmatfree}
\end{eqnarray}
where 
\begin{equation}
W_F(a_{L_h-1},a_{L_h};b_{L_h-1},b_{L_h})=z^{b_{L_h-1}+b_{L_h}}.
\label{bw3}
\end{equation}
The corresponding transfer matrices for hard hexagons are obtained by
supplementing (\ref{bw1}) with 
\begin{equation}
W_{}(a_j,a_{j+1};b_j,b_{j+1})=0 ~~{\rm for}~~a_{j+1}b_{j}
=1.
\label{bw4}
\end{equation}

We will consider four types of boundary conditions.

The grand  partition function  for  $L_v\times L_h$ 
lattices with periodic boundary conditions in both the 
$L_v$ and $L_h$ directions is given in terms of $T_C$ as  
\begin{eqnarray}
\label{transper}
\hspace{-0.9in}&& \qquad \quad  \quad  \quad 
Z^{CC}_{L_v,L_h}(z) \,\,\,\, = \,\,\,\,  {\rm Tr}\,\, T_C^{L_v}(z;L_h).
\end{eqnarray}

For free boundary conditions in the horizontal direction 
and periodic boundary conditions the vertical direction the partition
function is obtained from $T_F$ as
\begin{eqnarray}
\label{transcf}
\hspace{-0.9in}&& \qquad \quad  \quad  \quad 
Z^{CF}_{L_v,L_h}(z) \,\,\,\, = \,\,\,\,  {\rm Tr}\,\, T_F^{L_v}(z;L_h).
\end{eqnarray}

For periodic boundary conditions in the horizontal direction 
and free  boundary conditions the vertical direction the partition
function is obtained from $T_C$ as
\begin{eqnarray} 
\label{transfree}
\hspace{-0.9in}&&  
Z^{FC}_{L_v,L_h}(z)  = \langle {\bf v}_B|T_C^{L_v-1}(z;L_h)|{\bf v}'_B\rangle, 
\end{eqnarray}
where ${\bf v}_B$ and ${\bf v}'_B$ are suitable vectors 
for the boundary conditions on rows $1$ and $L_v$. 
For the transfer matrix (\ref{tmat}) with Boltzmann weights given 
by the asymmetrical form (\ref{bw1}), (\ref{bw2}) 
the components of the vectors 
 ${\bf v}_B$ and ${\bf v}'_B$ for free boundary conditions are
\begin{eqnarray} 
\hspace{-0.9in}&&  
{\bf v}_B(a_1,a_2,\cdots, a_{L_h})= \prod_{j=1}^{L_h}\,
z^{a_j},~~~~~{\bf v}'_B(b_1,b_2,\, \cdots,\, b_{L_h})=1.
\label{vector}
\end{eqnarray}
These vectors are invariant under translation and reflection.
 
For free boundary conditions in both directions
\begin{eqnarray} 
\label{freefree}
\hspace{-0.9in}&&  
Z^{FF}_{L_v,L_h}(z)  = \langle {\bf v}_B|T_F^{L_v-1}(z;L_h)|{\bf v}'_B\rangle, 
\end{eqnarray}

When the transfer matrix is diagonalizable
(\ref{transper})-(\ref{freefree}) may be written in terms of the
eigenvalues $\lambda_k$  and eigenvectors
 ${\bf v}_k$ of the transfer matrix
\begin{eqnarray}
\hspace{-0.9in}&&   
Z^{CC}_{L_v,L_h}(z) = \sum_k  \lambda_{k;C}^{L_v}(z;L_h),\label{eigent}\\ 
\hspace{-0.9in}&&   
Z^{CF}_{L_v,L_h}(z) = \sum_k  \lambda_{k:F}^{L_v}(z;L_h),\label{eigencf}\\ 
\hspace{-0.9in}&&   
Z^{FC}_{L_v,L_h}(z) =  \sum_k\lambda_{k;C}^{L_v-1}(z;L_h)\cdot
d_{C,k}~~ {\rm where}~~    
d_{C,k}\, =(\bf{v}_B\cdot \bf{v}_{C,k})(\bf{v}_{C,k} \cdot \bf{v}'_B),
\label{eigenfc}\\
\hspace{-0.9in}&&   Z^{FF}_{L_v,L_h}(z)=
\sum_k\lambda_{k;F}^{L_v-1}(z;L_h)\cdot  d_{F,k}~~
{\rm where}~~
d_{F,k}=  (\bf{v}_B\cdot \bf{v}_{F,k})(\bf{v}_{F,k} \cdot \bf{v}'_B).
\label{eigenff}
\end{eqnarray}

For hard squares and hard hexagons the transfer matrices $T_C(z;L_h)$
are invariant under translations and reflections and thus
momentum $P$ and parity $\pm$ are good quantum numbers. Furthermore
the boundary vectors ${\bf  v}_B$ and ${\bf v'}_B$ of (\ref{vector})
are invariant under translation and reflection, and consequently
$d_{C,k}=0$ unless the eigenvectors ${\bf v}_k$ lie in the positive
parity sector $P=0^+$.

For hard squares the matrix $T_F(z;L_h)$ is invariant under reflection
so the eigenvectors in the scalar products are restricted to positive
parity states. However for hard hexagons $T_F(z;L_h)$ is not invariant
under reflection and all eigenvectors will contribute to (\ref{eigenff}).

Note that partition function zeros for
all four boundary conditions have previously been studied for
antiferromagnetic Potts models \cite{ss}-\cite{JS5}. In that
case the relations to transfer matrix eigenvalues were similar to
(\ref{eigenfc}),(\ref{eigenff}). However, with periodic 
boundary conditions along the
transfer direction the partition function was defined as a Markov
trace, and  (\ref{eigent}),(\ref{eigencf}) were replaced by 
expressions involving
non-trivial eigenvalue multiplicities \cite{RJ1,RJ2}.

\subsection{Integrability}

To compare integrable with non-integrable systems a definition of
integrability is required.

The notion of integrability originates in the discovery by Baxter 
 that the Ising model and the 6 and 8 vertex models, which have
 transfer matrices  that depend on several variables, have a one
 parameter subspace for which the transfer matrices with different
 parameters will commute if cyclic boundary conditions are
 imposed \cite{baxbook}. This global property of the transfer matrix
 follows from a local property of the Boltzmann weights used to
 construct the transfer matrix, known as the star-triangle or the Yang-Baxter 
equation.

The hard hexagon model has only one
 parameter, the fugacity, but is also referred to as integrable because
Baxter \cite{baxterhh,baxbook} found that it may be realized
 as a special case of the model of
 hard squares with diagonal interactions which does have a one
 parameter family of commuting transfer matrices with cylindrical
 boundary conditions.

This concept of integrability has been generalized to transfer matrices
with boundary conditions which are not cylindrical if special boundary
conditions are imposed which satisfy a generalization of the
Yang-Baxter equation \cite{cher,skl} known as 
the boundary Yang-Baxter equation. This has been investigated for models  
closely related to hard hexagons \cite{behpaul,ahn} 
but the specialization
to hard hexagons with free boundary conditions has apparently not been made.

\subsection{The physical free energy}
\label{physicalfree}

For thermodynamics we are concerned with the limit
$L_v,~L_h\rightarrow \infty$, and in the physical region where
$z$ is real and positive the partition function per site $\kappa (z)$, the physical free
energy $F(z)$ and the density $\rho(z)$ are  defined as limits 
of the finite size grand partition function as
\begin{equation} 
\kappa(z) =
 \lim_{L_v,L_h\rightarrow \infty}Z_{L_v,L_h}(z)^{1/L_vL_h},
\label{free}
\end{equation} 
\begin{eqnarray}  
-F(z)/k_BT= 
 \lim_{L_v,L_h\rightarrow \infty}(L_vL_h)^{-1} \cdot \, \ln Z_{L_v,L_h}(z)
\label{free}
\end{eqnarray} 
and
\begin{equation}
\rho(z)=-z\frac{d}{dz}F(z).
\end{equation}
This limit must be independent of the
boundary conditions and aspect ratio $0\, <\,  L_v/L_h\, <\, \infty$ 
for thermodynamics to be valid. The free energy vanishes and is
analytic at $z=0$. For hard hexagons as $z\rightarrow \infty$
\begin{equation} 
F(z)/k_BT=\frac{1}{3}\ln z+{\tilde F}_{HH}(z)~~{\rm
  and}~~\rho(z)\rightarrow \frac{1}{3} 
\end{equation}
and for hard squares
\begin{equation}
F(z)/k_BT=\frac{1}{2}\ln z+{\tilde F}_{HS}(z)~~
{\rm and}~~\rho\rightarrow \frac{1}{2},
\end{equation}
where ${\tilde F}_{HH}(z)$ and ${\tilde F}_{HS}(z)$ are analytic at $z\rightarrow \infty$.
From this formulation series expansions of the free energy 
about both $z=0$ and $1/z=0$ are derived. The partition function per
site, physical free energy and density  for
$0\leq z \leq z_c$ and $z_c\leq z \leq \infty$ are different functions
which are not related to each other by analytic continuation around
the singularity at $z_c$. For hard hexagons the density
for both the low and the high density regime may be continued to the
full $z$ plane which for low density is cut from $-\infty \leq z \leq
z_{d;hh}$ and $z_{c;hh}\leq z \leq \infty$ and 
for high density cut from $z_{d;hh}\leq z
\leq z_{c;hh}$. Indeed, both the low and high density partition
functions per site and the density for hard hexagons are algebraic 
functions 
\cite{joyce,assis} and thus have analytic 
continuations even beyond the cuts in the $z$ plane.

To study the possibility of analytic continuation for hard squares 
of the physical partition function per site and density from the 
positive $z$ axis 
into the complex $z$ plane we consider both the formulation in terms 
of the transfer matrix and the zeros of the partition function.

\subsection{Analyticity and transfer matrix eigenvalues}
\label{analyticity}
 
For $0< z < \infty$  
all matrix elements of the transfer matrices  are positive   
so the Perron-Frobenious theorem  guarantees that
the largest eigenvalue $\lambda_{\rm max}$ is positive and the
corresponding eigenvector has all positive entries. Thus for all cases
\begin{eqnarray} 
\hspace{-0.9in}&& \qquad \quad  \quad  \quad 
\lim_{L_v\rightarrow \infty}\, L_v^{-1} \cdot \, \ln Z_{L_v,L_h}(z)
\,\, \,  =\, \,\, \,  \ln \lambda_{\rm max}(z;L_h)
\end{eqnarray}
and thus the free energy is
\begin{equation}
-F/k_BT=\lim_{L_h\rightarrow \infty}L_h^{-1}\ln \lambda_{\rm max}(z;L_h).
\label{transmax}
\end{equation}

Furthermore the cylindrical transfer
matrices for both squares and hexagons have  translation and 
reflection invariance. Therefore 
the eigenvalues of the lattice translation operator 
are $e^{iP}$ where $P$, the total 
momentum, has the values $2\pi n/L_h$, and the eigenvalues of the
reflection operator are $\pm 1$.  Each  transfer matrix eigenvalue 
has a definite
value of $P$ and parity  and $\lambda_{\rm max}$ has $P=0^+$ (where
$+$ indicates the reflection eigenvalue). Therefore for $0\leq z \leq
\infty$ the eigenvalue  $\lambda_{\rm max}$ of the transfer matrix
$T_C$ is the eigenvalue of an eigenvector in the sector $P=0^+$. 

To obtain the analytic continuation of the
density from the positive $z$ axis into the complex $z$
plane we need to continue the limit as $L_h\rightarrow \infty$ of the 
eigenvalue  with $P=0^+$ which is maximum on the positive axis.
However, the analytic continuation of $\lambda_{\max}$ off of the
segment $0\leq z \leq \infty$ will not, of course, have the largest
modulus in the entire complex $z$ plane.   
The analytic continuation of $\lambda_{\rm max}$ will be maximum only  
as long as it
has the largest modulus of all the eigenvalues and ceases to be
maximum when $z$ crosses an equimodular curve where the moduli of two (or more)
eigenvalues are the same. It is thus of
importance to determine the thermodynamic limit of the equimodular
curves of the largest eigenvalues of the transfer matrix.
In the thermodynamic limit the regions of $0\leq z \leq z_c$ and
$z_c\leq z \leq \infty$ are separated by one or more of these equimodular
curves. In \cite{assis} it was seen that for hard hexagons with 
finite $L_h$ the equimodular curves separate the $z$ plane into 
several regions. However, because the eigenvectors  with different 
momentum and parity lie in different subspaces only the eigenvalues
corresponding to eigenvectors with $P=0^+$ can affect the analytic
continuation of the density.

For the hard square transfer matrix with free boundary conditions, 
$T_F(z;L_h)$, the eigenvalue  $\lambda_{\mathrm{max}}$
will lie in the positive parity sector for positive $z$ and the
analytic continuation off the positive real axis will be constrained
to eigenvalues in the positive parity sector. For hard hexagons, where 
$T_F(z;L_h)$ is not reflection symmetric, $\lambda_{\mathrm{max}}$ is
not constrained to lie in a restricted sub-space.

It is thus clear from the  formulation of the physical free energy and
the density in
terms of the transfer matrix  that the process of
analytic continuation off of the positive $z$ axis and the taking 
of the thermodynamic limit do not commute. In the thermodynamic limit
it is not even obvious that for a non-integrable model an 
analytic continuation through the
limiting position of the equimodular curves is possible.  

\subsection{Analyticity and partition function zeros}

The considerations of analytic continuation in terms of partition function
zeros is slightly different because by definition polynomials are
single valued. However, once the thermodynamic limit is taken the
limiting locations of the zeros will in general divide the complex $z$
plane into disconnected zero free regions. For hard squares and hard
hexagons the
physical segments $0\leq z <z_c$ and $z_c<z<\infty$ lie in two
separate zero free regions. The density is  uniquely 
continuable into the zero free region and in these regions the free
energy will be independent of boundary conditions and aspect
ratio. For hard hexagons the density for both the low and high density
cases are further continuable beyond the zero free region 
into the respective cut planes of section \ref{physicalfree}. 
However, for hard squares  there is no 
guarantee that further continuation outside the zero free  regions is 
possible.

\subsection{Relation of zeros to equimodular curves}
\label{zerostophase}
For finite lattices the
partition function zeros can be obtained for $Z^{CC}_{L_v,L_h}(z)$ and
$Z^{CF}_{L_v,L_h}(z)$ from
(\ref{eigent}) and (\ref{eigencf}) if all eigenvalues are
known. For $Z^{FC}_{L_v,L_h}(z)$ and $Z^{FF}_{L_v,L_h}(z)$  
both the eigenvalues and eigenvectors are needed to obtain 
the zeros from (\ref{eigenfc}) and
(\ref{eigenff}).

The limiting case where
\begin{eqnarray}
\label{emod1}
\hspace{-0.9in}&& \qquad \quad  \quad \quad  \quad 
L_v \, \,\,  \rightarrow \,\,  \,\infty 
\quad \quad  \quad  {\rm with~ fixed} \quad  \quad  \quad  L_h, 
\end{eqnarray}
is presented in \cite{ss}-\cite{JS5},\cite{wood}-\cite{baxter1987}
with various boundary conditions 
 extending the work of
\cite{bkw1}-\cite{bkw3}. 
In this limit (\ref{emod1}) the partition function will 
have zeros when two or more maximum eigenvalues 
of $T(z;L_h)$ have equal moduli
\begin{eqnarray} 
\label{emod2}
\hspace{-0.9in}&& \qquad \quad  \quad  \quad   \quad 
|\lambda_1(z;L_h)| \, \,  =\, \, \,  |\lambda_2(z;L_h)|. 
\end{eqnarray} 
 
Consider first $Z^{CC}_{L_v,L_h}(z)$ and $Z^{CF}_{L_v.L_h}(z)$ where we see    
from (\ref{eigent}) and (\ref{eigencf}) that only eigenvalues are
needed. Thus, for these two cases, when only two largest
eigenvalues $\lambda_{1,2}$ need to be considered we may write
\begin{equation}
Z_{L_v,L_h}(z)=\lambda_1^{L_v}\left[1
+\left(\frac{\lambda_2}{\lambda_1}\right)^{L_v}+\cdots\right].
\label{tapprox}
\end{equation}
Then at values of $z$ where $|\lambda_1|=|\lambda_2|$ with
$\lambda_2/\lambda_1=e^{i\theta}$ we have for large $L_v$
\begin{equation}
Z_{L_v,L_h}(z)=\lambda_1^{L_v}[1+e^{i\theta L_v}+\cdots]
\end{equation}
and hence $Z_{L_v,L_h}(z)$ will have a zero close to this $z$ when
\begin{equation}
e^{i\theta L_v}=-1,
\label{thetalv}
\end{equation}
that is when
\begin{equation}
\theta L_v=(2n+1)\pi
\label{thetan}
\end{equation} 
with $n$ an integer. This relation becomes exact in the
limit $L_v\rightarrow \infty$. Calling $z_i$ and $z_{i+1}$ the values
of $z$ at two neighboring zeros on the equimodular curve we thus
obtain from (\ref{thetan})
\begin{equation}
\theta(z_{i+1})-\theta(z_{i})=2\pi/L_v.
\label{thetadiff}
\end{equation}

Let $s(z)$ be the arclength along an equimodular curve. Then the
derivative of $\theta(s(z))$ with respect to $s$ 
is defined as the limit of
\begin{equation}
\frac{\Delta\theta}{\Delta s}\equiv
\frac{\theta(s(z_{i+1}))-\theta(s(z_i))}{s(z_{i+1})-s(z_i)},
\label{deltadef}
\end{equation}
Thus, defining the density of roots on the equimodular curve as
\begin{equation}
D(s)=\lim_{L_v\rightarrow \infty}\frac{1}{L_v[s(z_{i+1})-s(z_i)]},
\end{equation}
we find from (\ref{thetadiff}) and (\ref{deltadef}) that for
$L_v\rightarrow \infty$ with $L_h$ fixed that the density of zeros on an
  equimodular curve is
\begin{equation}
\frac{d\theta(s)}{ds}=2\pi D(s).
\label{phidensity}
\end{equation}

For $Z^{FC}_{L_v,L_h}(z)$ and $Z^{FF}_{L_v,L_h}(z)$ from
(\ref{eigenfc}) and (\ref{eigenff}) we 
have instead of (\ref{tapprox})
\begin{equation}
Z_{L_v,L_h}(z)=\lambda_1^{L_v}d_1\left[1
+\left(\frac{\lambda_2}{\lambda_1}\right)^{L_v}\frac{d_2}{d_1}+\cdots\right],
\end{equation}
with
\begin{equation}
\frac{d_2}{d_1}=re^{i\psi},
\end{equation}
where in general $r\neq 1$. Thus writing
 \begin{equation}
\frac{\lambda_2}{\lambda_1}=\epsilon e^{i\theta},
\end{equation}
the condition for a zero in the limit $L_v\rightarrow \infty$
which generalizes (\ref{thetalv}) is
\begin{equation}
\epsilon^{L_v}e^{i\theta L_v}re^{i\psi}=-1,
\end{equation}
from which we obtain
\begin{eqnarray}
&&\epsilon= r^{-1/L_v}=e^{-\ln r/L_v}\sim 1-\frac{\ln r}{L_v},\\
&&\theta L_v+\psi=(2n+1)\pi.
\end{eqnarray}
Thus as $L_v\rightarrow \infty$ the locus of zeros
  approaches the equimodular curve as $\ln r/L_v$ and the limiting
  density is still given by (\ref{phidensity}).



These considerations, however, are in general not sufficient for the
study of the thermodynamic limit where instead of (\ref{emod1}) we are
interested in the limit
\begin{equation}
L_v\rightarrow \infty,~~L_h\rightarrow \infty,~~~{\rm with~~
  fixed}~~L_v/L_h
\label{thermolimit}
\end{equation}
and  the physical free energy must be independent of the 
aspect ratio $L_v/L_h$.  

To study the limit (\ref{thermolimit}) there are several properties of the 
dependence of the equimodular curves on $L_h$   which need to be
considered:
\begin{enumerate}

\item The derivative of the phase  $\theta(s)$  on a curve can vanish 
as $L_h\rightarrow \infty$ on some portions of the curve;

\item The number of equimodular curves can diverge as $L_h\rightarrow \infty$ 
and there can be regions in the $z$ plane where they become dense;

\item The length of an equimodular curve can vanish as $L_h\rightarrow
   \infty$.

\end{enumerate}
The first of these properties is illustrated for hard hexagons in
\cite{assis}. The second and third properties have been observed for
antiferromagnetic Potts models in \cite{CJSS}. 

We will see that all three phenomena are present for
hard squares. The roots of the $L\times L$ partition function in the
limit $L\rightarrow \infty$ converge to lie on the 
 $L_h \rightarrow \infty$  limit of the equimodular curves.

\section{Global comparisons of squares and hexagons}
\label{compare}

In \cite{assis} we computed for hard hexagons the
zeros for  $L\times L$ lattices of $Z^{CC}_{L,L}(z)$ for toroidal boundary
conditions, and for cylindrical boundary conditions where
\begin{equation}
Z^{FC}_{L,L}(z)=Z^{CF}_{L,L}(z).
\end{equation}
We also computed the equimodular curves for both  the full transfer 
matrix $T_C(z;L_h)$ relevant to $Z^{CC}_{L_v,L_h}(z)$ and the restriction
to the subspace $P=0^+$ relevant for $Z^{FC}_{L_v,L_h}(z)$.
In this paper we compute the same quantities for
hard squares and compare them with the results of \cite{assis}. 
We also compute the equimodular curves for $T_F(z;L_h)$ relevant for 
$Z^{CF}_{L_v,L_h}(z)$ and $Z^{FF}_{L_v,L_h}(z)$. For
hard hexagons we restricted attention to $L_v,~L_h$  multiples of three
which is commensurate with hexagonal ordering. Similarly for hard
squares we restrict attention here to $L_v,~L_h$ even to be
commensurate with square ordering.

\subsection{Comparisons of partition function zeros}

\begin{figure}[h!]

\begin{center}
\hspace{0cm} \mbox{
\begin{picture}(300,140)
\put(0,0){\includegraphics[width=5cm]{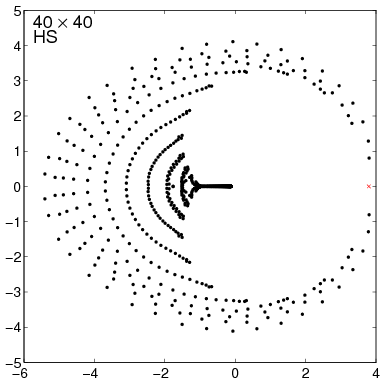}}
\put(150,0){\includegraphics[width=5cm]{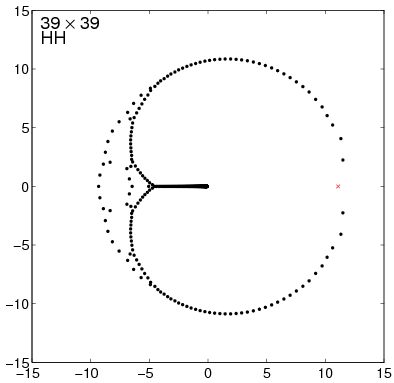}}
\end{picture}
}
\end{center}

\caption{Comparison in the complex fugacity plane $z$ of the 
zeros of the partition function $Z^{FC}_{L,L}(z)=Z^{CF}_{L,L}(z)$ with 
cylindrical boundary conditions of hard squares on the $40\times 40$ lattice 
on the left to hard hexagons  on the $39\times 39$ lattice on the
right. The location of $z_c$ and $z_{c;hh}$ is indicated by a cross.}
\label{fig:cf}
\end{figure}

\begin{figure}[h!]

\begin{center}
\hspace{0cm} \mbox{
\begin{picture}(300,140)
\put(0,0){\includegraphics[width=5cm]{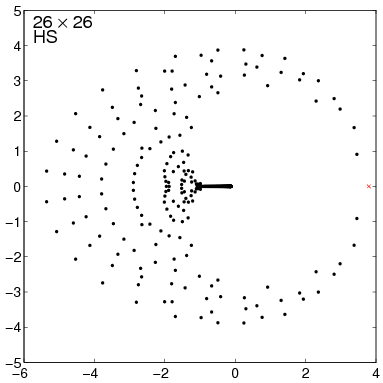}}
\put(150,0){\includegraphics[width=5cm]{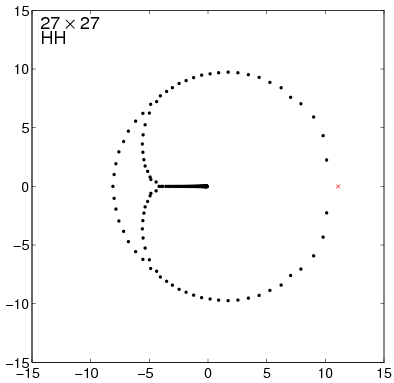}}
\end{picture}
}
\end{center}

\caption{Comparison in the complex fugacity plane $z$ of the 
zeros of the partition function $Z^{CC}_{L,L}(z)$ with 
toroidal boundary conditions of hard squares on the $26\times 26$ lattice 
on the left to hard hexagons on the $27\times 27$ lattice on the
right.The location of $z_c$ and $z_{c;hh}$ is indicated by a cross.}
\label{fig:torus}
\end{figure}

\begin{figure}[h!]

\begin{center}
\hspace{0cm} \mbox{
\begin{picture}(300,140)
\put(0,0){\includegraphics[width=5cm]{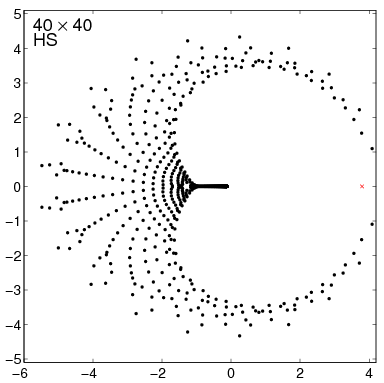}}
\put(150,0){\includegraphics[width=5cm]{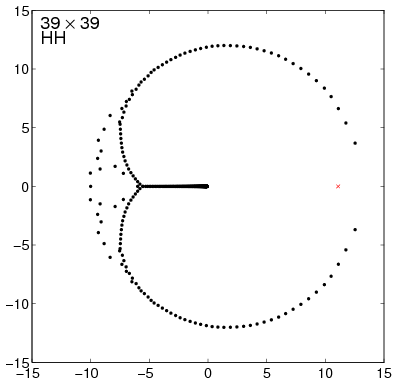}}
\end{picture}
}
\end{center}

\caption{Comparison in the complex fugacity plane $z$ of the 
zeros of the partition function $Z^{FF}_{L,L}(z)$ with 
free boundary conditions  of hard squares on the $40\times 40$ lattice 
on the left to hard hexagons on the $39\times 39$ lattice on the
right. The location of $z_c$ and $z_{c;hh}$ is indicated by a cross.}
\label{fig:free}
\end{figure}

We have computed zeros of the hard square partition function 
in the complex fugacity $z$ plane  for  $L\times L$ lattices
with cylindrical  and free boundary conditions for
$L\leq 40$  and for toroidal boundary conditions for $L\leq 26$ using
the methods of \cite{assis}. In figure \ref{fig:cf} we compare
partition function zeros for cylindrical boundary conditions 
of hard squares on the $40\times 40$ lattice with hard hexagons on the
$39\times 39$ lattice and in figure \ref{fig:free} the comparison is
made for free boundary conditions. In figure \ref{fig:torus} we 
  compare for toroidal boundary conditions hard squares on the
  $26\times 26$ lattice with hard hexagons on the $27\times 27$ lattice.

For both hard squares and hard hexagons there is a line of zeros on
the negative real axis ending at $z_d$ and $z_{d;hh}$,
respectively. The ratio of real roots to complex roots for hard
squares is roughly 1/2:1/2 while for hard hexagons the ratio is
roughly 2/3:1/3.

The most obvious difference between hard squares and hard hexagons  in 
figures \ref{fig:cf}-\ref{fig:free}  is that the zeros of hard
squares are seen to lie in an area instead of being confined to
a few well defined curves as is seen for hard hexagons. 

For cylindrical boundary conditions the filling up of this area 
proceeds in a remarkably regular fashion. 

For the lattices $4N\times 4N$ there are $N-1$ outer arcs
each of $4N$ points, then there is a  narrow arclike area with close to $4N$
zeros and finally there is 
an inner structure that is  connected to $z=-1$. For the innermost of the
$N-1$ arcs the zeros appear in well defined pairs.

For lattices $(4N+2)\times (4N+2)$ there are $N-1$ outer arcs each of
$4N+2$ points, then a narrow arclike area which has close to $4N+2$
zeros and finally an inner structure that is connected to $z=-1$.

For all boundary conditions the zeros of hard squares
appear to converge in the $L\rightarrow
\infty$ limit to a wedge which hits the positive $z$ axis at
$z_c$. This is distinctly different from the behavior of hard hexagons
 where the zeros appear to approach $z_{c;hh}$ on a well defined one  
dimensional arc.

\begin{figure}[!htp]

\begin{center}
\hspace{0cm} \mbox{
\begin{picture}(300,300)
\put(0,0){\includegraphics[width=10cm]{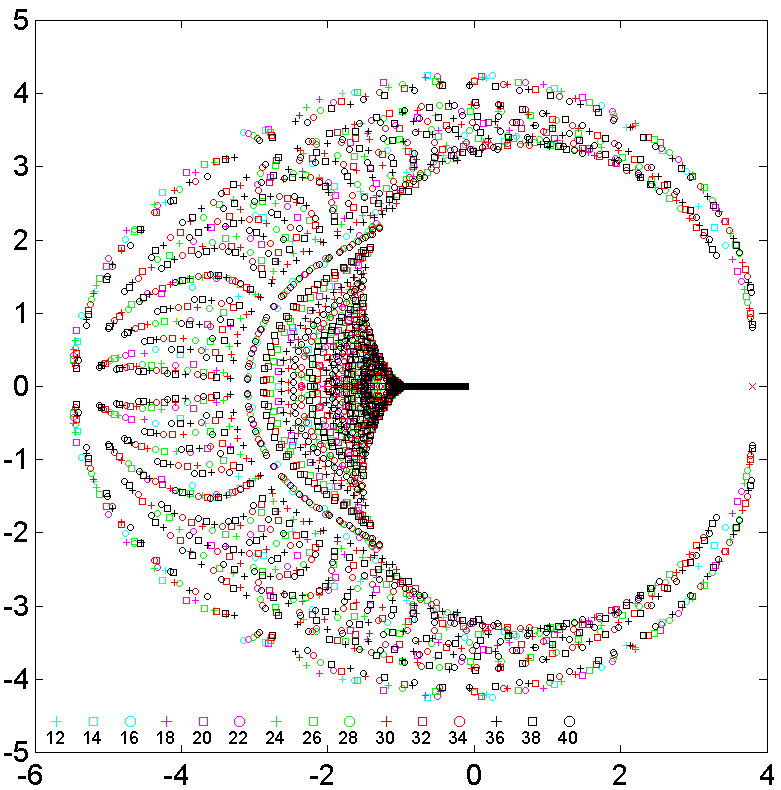}}
\end{picture}
}
\end{center}
\caption{Combined plot of hard square zeros of
 $Z^{CF}_{L,L}(z)=Z^{FC}_{L,L}(z)$ for  the $L \times L$ lattice
with cylindrical boundary conditions for $12\leq L \leq 40$. We
exhibit a mod six effect by plotting $L=6n+4$ as circles, $L=6n+2$ as
boxes and $L=6n$ as crosses, The values of $L_h$ are shown in the
 different colors indicated in the legend. It is to be noticed that there is a
distinguished curve  where only points $L=6n+4$ lie. The location of
 $z_c$ is indicated by a cross.}
\label{fig:hsall}
\end{figure}

In figure \ref{fig:hsall} we illustrate the dependence on $L$ of the hard
square zeros of $Z^{FC}_{L,L}(z)=Z^{CF}_{L,L}(z)$ of 
the $L\times L$ lattice by giving a combined plot 
of all the zeros for $12\leq L \leq 40$. This reveals that the three
cases of $L=6n+4,~6n+2$ and $6n$ approach the common limit in three
separate ways. There is one  well
defined curve whose position does not depend on $L$ which consists 
only of the points of $L=6n+4$ lattices. 

In table \ref{tab:endpoints} we list the value of the zero closest to the
three endpoints $z_c,~z_d$ and $-1$ for the $L\times L$ cylindrical 
lattices with
$24\leq L\leq 40$. We also list the number $N_L$ of zeroes  in 
$-1\leq z \leq z_d$ plus  the number of zeroes
$z< -1$. For $L=40$ we note that ${\rm Re}[z_c(40)]>z_c$ whereas for
$L\leq 38$ we have   ${\rm Re}[z_c(L)]<z_c$. This behavior of $z_c(L)$
in relation to $z_c$ is similar to what is seen for hard hexagons
in table 5 of \cite{assis} where ${\rm Re }[z_c(L)]>z_c$ for $L\geq 21$ and only
starts to approach $z_c$ from the right for $L=36$.


\begin{table}[h]
\begin{center}
\begin{tabular}{|l|l|l|l|l|}\hline
$L$& $z_c(L)$&$z_d(L)$&$z_{-1}(L)$&$N_L$\\ \hline
$24$&$3.690334\pm i 1.324109$&$-0.119976$&$-0.956723$&$128+0$\\
$26$&$3.718433\pm i 1.226238$&$-0.119871$&$-0.979835$&$153+5$\\
$28$&$3.739986\pm i 1.141529$&$-0.119788$&$-0.986589$&$176+0$\\
$30$&$3.756751\pm i 1.067554$&$-0.119723$&$-0.991656$&$201+5$\\
$32$&$3.769947\pm i 1.002431$&$-0.119671$&$-0.992168$&$231+1$\\
$34$&$3.780438\pm i 0.944686$&$-0.119628$&$-0.989045$&$259+9$\\
$36$&$3.788852\pm i 0.893150$&$-0.119592$&$-0.976523$&$288+0$\\
$38$&$3.795647\pm i 0.846884$&$-0.119563$&$-0.994325$&$325+9$\\
$40$&$3.801169\pm i 0.805129$&$-0.119538$&$-0.991673$&$358+0$\\
$\infty$&$3.796255$&$-0.119338$&$-1$\\
\hline
\end{tabular}
\end{center}
\caption{The endpoints $z_c(L),~z_d(L)$ and $z_{-1}(L)$ for
  the $L\times L$ cylindrical lattices with $24\leq L\leq 40$. The number of
  zeros $N_L$ on the segment $-1\leq z \leq z_d$ as well as the very 
small number of points $z\leq -1$ which do not contribute to the density.}
\label{tab:endpoints}
\end{table}

\subsection{Comparisons of  equimodular curves with partition zeros}

We have computed equimodular curves for the hard square 
transfer matrix $T_C(z;L_h)$ in  the sector $P=0^+$ for even
$L_h\leq 26$  and for the full transfer matrix for $L_h\leq 18$.  
For hard squares we have computed the equimodular curves for the full
$T_F(z;L_h)$ and the restriction to the positive parity sector for
$L_h\leq 16$. For 
hard hexagons  the equimodular curves of $T_C(z;L_h)$ were computed in
\cite{assis} for $L_h\leq 21$ and in the sector $P=0^+$ for $L_h\leq
30$. Equimodular curves for the hard hexagon transfer matrix
$T_F(z;L_h)$ are computed here for $L_h\leq 21$.

\begin{figure}[h!]

\begin{center}
\hspace{0cm} \mbox{
\begin{picture}(400,400)
\put(0,200){\includegraphics[width=7cm]{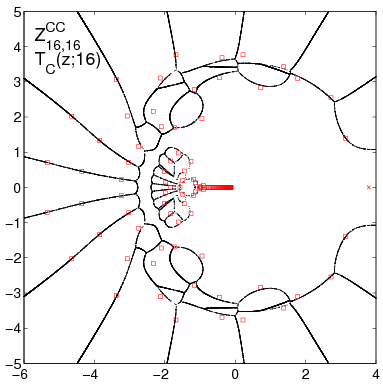}}
\put(200,200){\includegraphics[width=7cm]{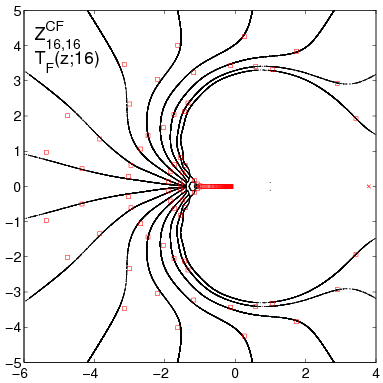}}
\put(0,0){\includegraphics[width=7cm]{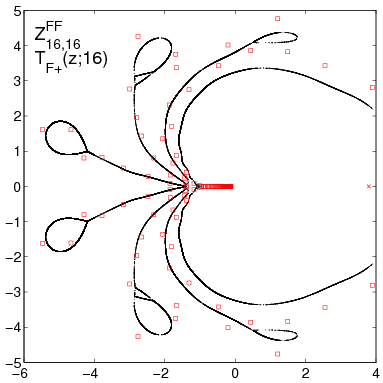}}
\put(200,0){\includegraphics[width=7cm]{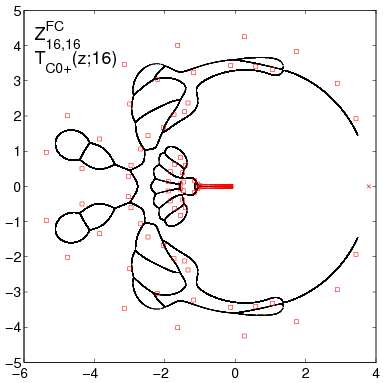}}
\end{picture}
}
\end{center}

\caption{Comparison for hard squares of the three types of zeros
  and the 4 types of equimodular curves. Clockwise from the upper
left we have for $L=16$:  $Z^{CC}_{L,L}(z)$ with $T_C(z;L)$, 
$Z^{CF}_{L,L}(z)$ with
  $T_F(z;L)$, $Z^{FC}_{L,L}(z)$ with $T_C(z;L)$ restricted to $P=0^+$ and
  $Z^{FF}_{L,L}(z)$ with $T_{F}(z;L)$ restricted to positive parity. We note that the zeros of
  $Z^{FC}_{L,L}(z)$ and $Z^{CF}_{L,L}(z)$ are identical even though
  the equimodular curves are very different. The location of $z_c$ is
  indicated by a cross.}
\label{fig:evzeroshs}
\end{figure}

\begin{figure}[h!]

\begin{center}
\hspace{0cm} \mbox{
\begin{picture}(400,400)
\put(0,200){\includegraphics[width=7cm]{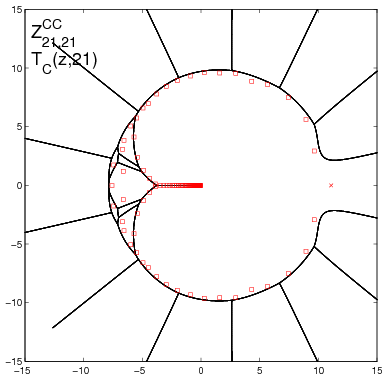}}
\put(200,200){\includegraphics[width=7cm]{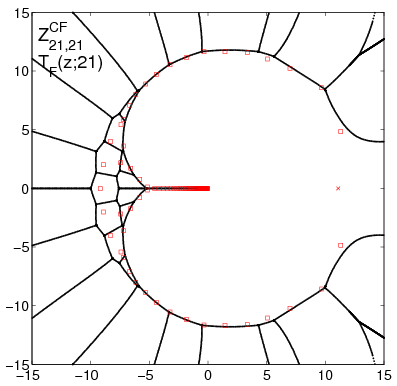}}
\put(0,0){\includegraphics[width=7cm]{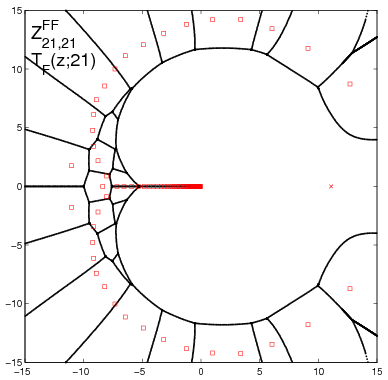}}
\put(200,0){\includegraphics[width=7cm]{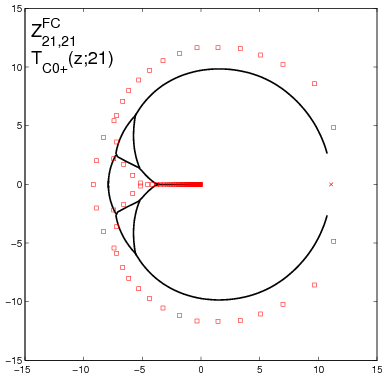}}
\end{picture}
}
\end{center}
\caption{Comparison for hard hexagons of the three types of zeros
  and the 3 types of equimodular curves. Clockwise from the upper
left we have for $L=21$:  $Z^{CC}_{L,L}(z)$ with $T_C(z;L)$, 
$Z^{CF}_{L,L}(z)$ with
  $T_F(z;L)$, $Z^{FC}_{L,L}(z)$ with $T_C(z;L)$ restricted to $P=0^+$ and
  $Z^{FF}_{L,L}(z)$ with $T_{F}(z;L)$. We note that the zeros of
  $Z^{FC}_{L,L}(z)$ and $Z^{CF}_{L,L}(z)$ are identical even though
  the equimodular curves are very different. The location of $z_{c;hh}$ is
  indicated by a cross.}
\label{fig:evzeroshh}
\end{figure}

In figure \ref{fig:evzeroshs} we plot the equimodular curves and zeros
for hard squares. This is to be compared with the similar plot for
hard hexagons in figure \ref{fig:evzeroshh}. In both cases we note that
the zeros for $Z^{FC}_{L,L}(z)$ and $Z^{CF}_{L,L}$ are identical while
the corresponding equimodular curves  are
different.

\clearpage

\clearpage

The equimodular curves of hard squares are strikingly different from
those of hard hexagons for all cases considered. The hard hexagon plots
consist of a few well defined sets of curves which, with the exception that  
the curves for $P=0^+$ do not have rays extending to infinity, are
qualitatively very similar for all four cases. For hard
squares, on the other hand, the four different plots are
qualitatively different from each other and are far more complicated
than those for hard hexagons. 
 
\begin{figure}[!h]

\begin{center}
\hspace{0cm} \mbox{
\begin{picture}(400,540)
\put(0,360){\includegraphics[width=6cm]{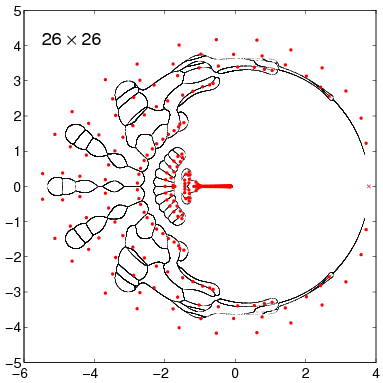}} 
\put(200,360){\includegraphics[width=6cm]{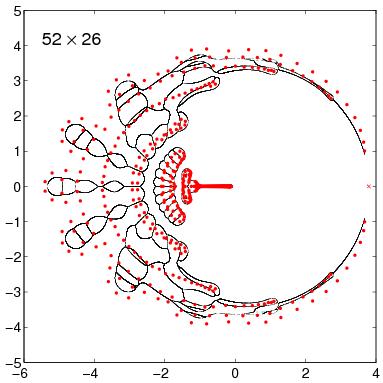}}
\put(0,180){\includegraphics[width=6cm]{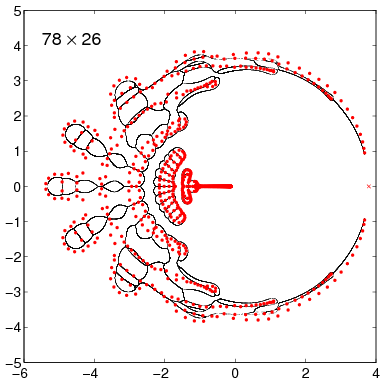}}
\put(200,180){\includegraphics[width=6cm]{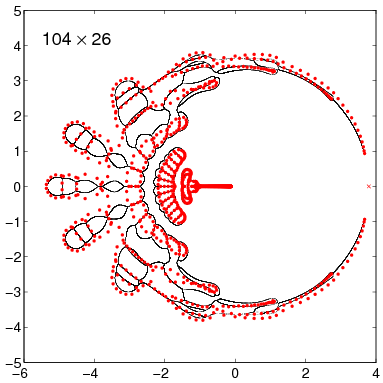}}
\put(0,0){\includegraphics[width=6cm]{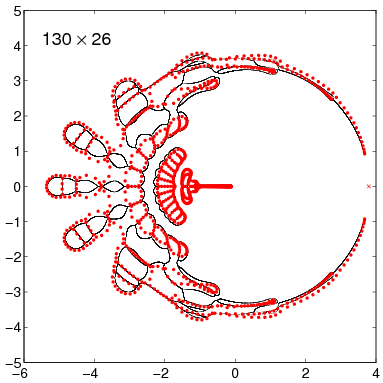}}
\put(200,0){\includegraphics[width=6cm]{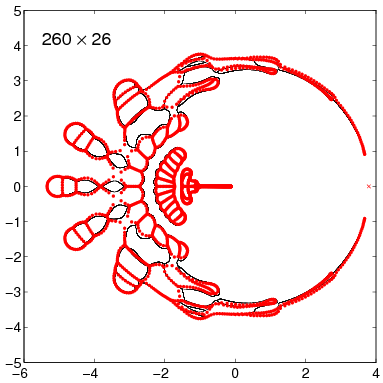}}
\end{picture}
}
\end{center}

\caption{Plots in the complex fugacity $z$-plane of the 
zeros of the partition function  $Z^{FC}_{L_v,L_h}(z)$ of hard squares 
for $L_v\times 26$
lattices with cylindrical boundary conditions (in red) compared with
the $P=0^+$ equimodular curves of $T_C(z;26)$ (in black). The location
of $z_c$ is indicated by a cross.}
\label{fig:anisozeros}
\end{figure}

The cylinder partition function  $Z^{FC}_{L,L}(z)=Z^{CF}_{L,L}(z)$
allows a direct comparison between the
equimodular curves of $T_F(z)$ and $T_{C0^+}(z)$ in figures 
\ref{fig:evzeroshs} and \ref{fig:evzeroshh}, since both transfer
matrices can be used to construct the same partition function. For
both hard squares and hard hexagons these figures show  that the zeros of the
$L\times L$ cylindrical partition function lie much closer to the equimodular
curves of $T_F(z)$ rather than $T_{C0^+}$. It is only for much larger
aspect ratios that the cylinder zeros lie close to the $T_{C0^+}$
equimodular curves, as can be seen, for example, in 
figure~\ref{fig:anisozeros}, where we plot the 
hard square $Z^{FC}_{26n,26}(z)$ roots for $n=1,~2,~3,~4,~5,~10$ along with the
$L_h=26$ equimodular curves of $T_{C0^+}(z)$.

For hard squares, the arclike structures noted above for figure \ref{fig:cf} 
are in remarkable agreement with the $T_F(z)$ curves which
originate near $z=-1$ and extend to infinity. There are $L_h$ such
equimodular curves which is exactly the number of points seen above to
lie on each of the arclike structures of zeros.

For hard squares both $T_C(z;L_h)$ and
$T_F(z;L_h)$ shown in figure \ref{fig:evzeroshs} have equimodular curves
which extend out to $|z|=\infty$. In \ref{appc} we present an
analytical argument that both the
$T_C(z;L_h)$  and $T_F(z;L_h)$ curves have $L_h$ branches going out 
to infinity at asymptotic angles $\arg z = \frac{(1+2k) \pi}{L_h}$ 
with $k=0,1,\ldots,L_h-1$. 

\begin{figure}[h!]

\begin{center}
\hspace{0cm} \mbox{
\begin{picture}(250,250)
\put(0,0){\includegraphics[width=11cm]{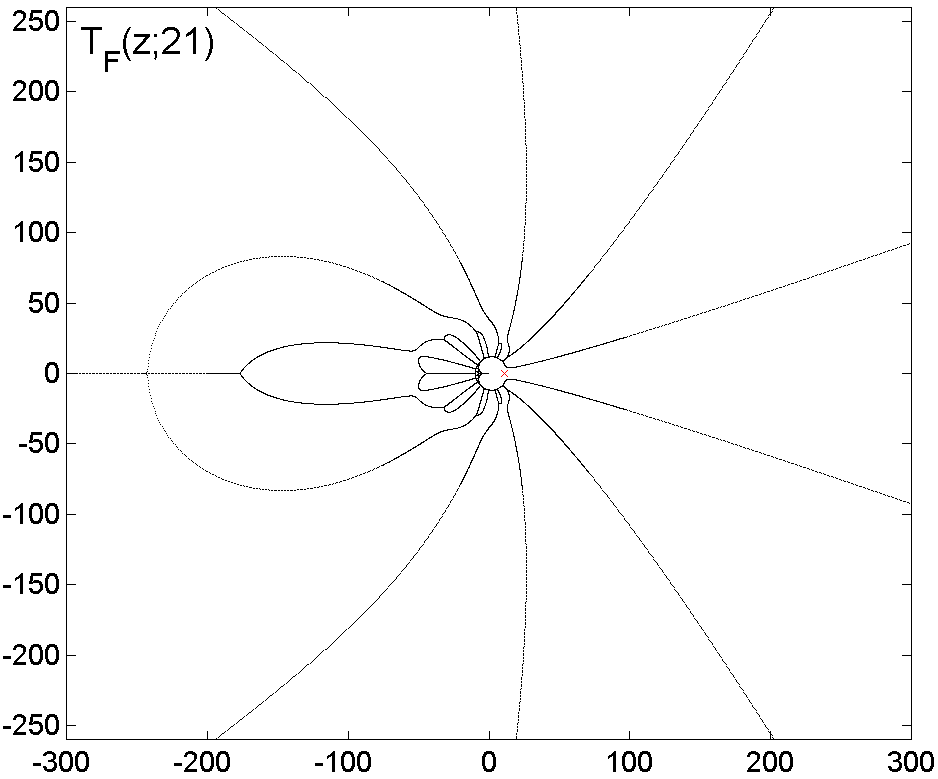}}
\end{picture}
}
\end{center}
\caption{Equimodular curves for the hard hexagon transfer matrix 
$T_F(z;L_h)$ for $L_h=21$ showing the complex structure which exists
for $|z|\geq 12$. The location of $z_c$ is indicated by a cross.}
\label{fig:hhlargez}
\end{figure}

For hard hexagons it was seen in
\cite{assis} that when $L_h\equiv 0~({\rm mod}~3)$  the curves for
$T_C(z;L_h)$ as illustrated in  figure \ref{fig:evzeroshh}  
have $2L_h/3$ rays extending to infinity which separate regions with
$P=0^+$ from regions with $\pm 2\pi/3$. However, for the hard hexagon 
matrix $T_F(z;L_h)$ it is evident in figure \ref{fig:evzeroshh} there is much
more structure in the curves which extend to infinity. This is shown on
a much larger scale in figure \ref{fig:hhlargez}.
This more complicated structure for the equimodular curves of 
$T_F(z;L_h)$ presumably results from the fact that for hard hexagons 
$T_F(z;L_h)$ is  neither translation nor reflection invariant. 

Just as for hard hexagons it is only possible for hard squares  
to identify an endpoint of an equimodular curve approaching $z_c$ 
for the transfer matrix
$T_C(z;L_h)$ in the $P=0^+$ sector. We give the location of the $z_c(L_h)$
and $z_d(L_h)$ endpoints for $P=0^+$
in table \ref{tab:curveend}.

\begin{figure}[!h]

\begin{center}
\hspace{0cm} \mbox{
\begin{picture}(400,180)
\put(0,0){\includegraphics[width=6cm]{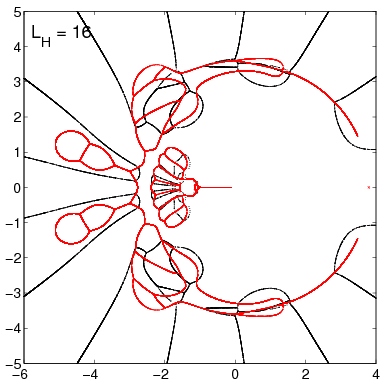}}
\put(200,0){\includegraphics[width=6cm]{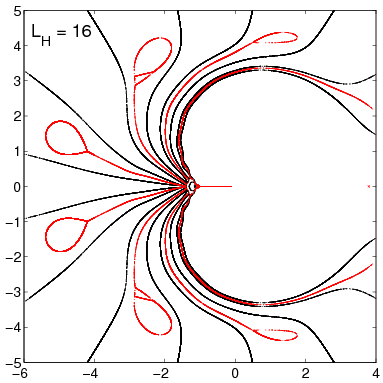}}
\end{picture}
}
\end{center}

\caption{On the left the comparison for hard squares  with  $L_h=16$ 
of the equimodular curves of
  $T_C(z;L_h)$ in black with the restriction to $P=0^+$ in red.
On the right the comparison for hard squares  with  $L_h=16$ 
of the equimodular curves of $T_F(z;L_h)$ in black with the
restriction to the positive parity sector in red. The location 
of $z_c$ is indicated by a cross.}
\label{fig:hstfcomp}
\end{figure}

For hard squares the transfer matrix $T_F(z;L_h)$ with free boundary
conditions is invariant under parity in contrast with hard hexagons
where there is no parity invariance. The maximum eigenvalue for hard
squares has positive parity and in figure \ref{fig:hstfcomp} we
compare for $L_h=16$ the equimodular curves of $T_F(z;L_h)$ with the 
restriction to positive parity. We also compare the equimodular curves
for $L_h=16$ of $T_C(z;L_h)$ and its restriction to $P=0^+$.

\clearpage

\clearpage

\begin{table}[h]
\begin{center}
\begin{tabular}{|r|l|l|}\hline
$L_h$& $z_c(L_h)~ {\rm endpoint}$&$z_d(L_h)~ {\rm endpoint}$\\ \hline
$4$&$-0.8806\pm i 3.4734$&$-0.1259$\\
$6$&$1.6406\pm i 3.2293$&$-0.1216$\\
$8$&$2.5571\pm i 2.6694$&$-0.1204$\\
$10$&$2.9955\pm i 2.2264$&$-0.1200$\\
$12$&$3.2374\pm i 1.8961$&$-0.1197$\\
$14$&$3.3845\pm i 1.6461$&$-0.1196$\\
$16$&$3.479\pm i 1.4547$&\\
$18$&$3.544\pm i 1.3032$&\\
$20$&$3.591\pm i 1.1780$&\\
$22$&$3.627\pm i 1.0722$&\\
$24$&$3.654\pm i 0.9841$&\\
$26$&$3.675\pm i 0.9117$&\\
$\infty$&$3.796255$&$-0.119338$\\
\hline
\end{tabular}
\end{center}
\caption{The endpoints of the equimodular curves of $T_C(z;L_h)$ with 
$P=0^+$ which approach $z_c$
  and $z_d$ as $L_h$ increases. For $L_h\leq 14$ the endpoints are
  computed from the vanishing of the discriminant of the
  characteristic polynomial and have been computed to 50 decimal
  places. For $L_h\geq 16$ they are determined numerically to 3
  decimal places and consequently the deviation from $z_d$ is too
  small to be accurately determined.} 

\label{tab:curveend}
\end{table}

\section{Comparisons on $-1\leq z \leq z_d$}
\label{negz}

A much more quantitative comparison of hard squares and hard hexagons 
can be given on the interval $-1\leq z\leq z_d$. We treat both transfer
matrix eigenvalues and partition function zeros.

\subsection{Transfer matrix eigenvalue gaps} 
\label{tmezm}

The eigenvalues of the transfer matrix $T_C(z;L_h)$ for 
hard hexagons for $P=0^+$ have two very remarkable properties discovered 
in \cite{assis} 

\begin{enumerate}

\item The characteristic polynomial of $T_C(z)$ in the sector $P=0^+$ for
   $L_h=9,~ 12,~15,~18$ factorizes into the product of two irreducible
   polynomials with integer coefficients.

\item  The roots of the discriminant of the characteristic polynomial 
which lie on the real axis for
$z<z_{d;hh}(L)$ all have multiplicity two for  $L_h\leq 18$.  
In particular on the 
negative real axis  the maximum eigenvalue is real only at isolated points.
We conjecture this is valid for all $L_h$.

\end{enumerate}

The hard hexagon transfer matrix $T_F(z;L_h)$ for $L_h=3,~6,~ 9$ also
has the remarkable property that all the roots of the resultant on the
interval $-1<z <z_d$ have multiplicity two. This is  very strong
evidence to support the conjecture that hard hexagons with free
boundary conditions in one direction and cyclic in the other direction
is obtained as a limit from a model which obeys the boundary
Yang-Baxter equation of \cite{cher,skl}.
 
Neither property i) nor ii) can be considered as being generic and
neither property holds for hard squares where there are small gaps 
in the equimodular curves where the maximum eigenvalues of 
both $T_c(z;L_h)$ and $T_F(z;L_h)$
are real and non-degenerate.
These gaps are caused by the collision of a complex conjugate 
pair of eigenvalues at the boundaries of the gaps.  
On $-1\leq z \leq z_d$ the maximum eigenvalue of $T_C(z;L_h)$ is in the
sector $P=0^+$. We have computed
these gaps numerically for $L_h\leq 20$ and more accurately 
from the discriminant of the characteristic polynomial for $L_h\leq
14$. We give these gaps in table \ref{tab:gaps1} for $L_h\leq 20.$ 
For $L_h\geq22$ most of the gaps are too small to actually
observe their width, but their locations can still be determined
numerically and are given in table \ref{tab:gaps2} for $22\leq L_h \leq
30$.

\begin{table}[!h]
\begin{center}
\begin{tabular}{|r|l|l|l|l|}\hline
$L_h$& $z_l(L_h)$&$z_r(L_h)$&gap&eigenvalue sign\\ \hline
$6$&$-0.52385422$&$-0.47481121$&$4.904301\times 10^{-2}$&$-$\\
$8$&$-0.30605227$&$-0.30360084$&$2.35243\times 10^{-3}$&$-$\\
$10$&$-0.23737268$&$-0.23720002$&$1.7266\times 10^{-4}$&$-$\\
&$-0.77929238$&$-0.73645527$&$4.283711\times 10^{-2}$&$+$\\
$12$&$-0.20401756$&$-0.20400239$&$1.517\times10^{-5}$&$-$\\
&$-0.49539291$&$-0.49352002$&$1.87289\times 10^{-3}$&$+$\\
$14$&$-0.18464415$&$-0.18464265$&$1.50\times 10^{-6}$&$-$\\
&$-0.37193269$&$-0.37180394$&$1.2875 \times 10^{-4}$&$+$\\
&$-0.92551046$&$-0.91949326$&$6.01721\times 10^{-3}$&$-$\\
$16$&$-0.17211444$&$-0.1721143$&$1.4\times 10^{-7}$&$-$\\
&$-0.305086$&$-0.305078$&$8\times 10^{-6}$&$+$\\
&$-0.64336$&$-0.64204$&$1.32\times 10^{-3}$&$-$\\
$18$&$-0.163389012$&$-0.163388998$&$1.4\times 10^{-8}$&$-$\\
&$-0.2643054$&$-0.2643045$&$9\times 10^{-7}$&$+$\\
&$-0.494482$&$-0.494388$&$9.4\times 10^{-5}$&$-$\\
$20$&$-0.156991031$&$-0.156991029$&$2\times 10^{-9}$&$-$\\
&$-0.23723539$&$-0.23723530$&$9\times 10^{-8}$&$+$\\
&$-0.404127$&$-0.494120$&$7\times 10^{-6}$&$-$\\
&$-0.7537$&$-0.7523$&$1.4\times 10^{-3}$&$+$\\
\hline
\end{tabular}
\end{center}
\caption{The gaps on the segment $-1\leq z \leq z_d$ where the maximum
  eigenvalue of the transfer matrix  $T_C(z;L_h)$ for hard squares 
on cylindrical chains  of length $L_h$ is real for $6\leq L_h \leq 20$.}
\label{tab:gaps1}
\end{table}

The gaps of $T_F(z;L_h)$ are not the same as those of $T_C(z;L_h)$. 
The gaps of
 $T_F(z;L_h)$ are given in table \ref{tab:gaps1f} where we see that
 with increasing $L_h$ they approach the gaps of $T_C(z;L_h)$
 of table \ref{tab:gaps1}.

\begin{figure}[!h]
\begin{center}
\hspace{0cm} \mbox{
\begin{picture}(350,150 )
\put(0,0){\includegraphics[width=6cm]{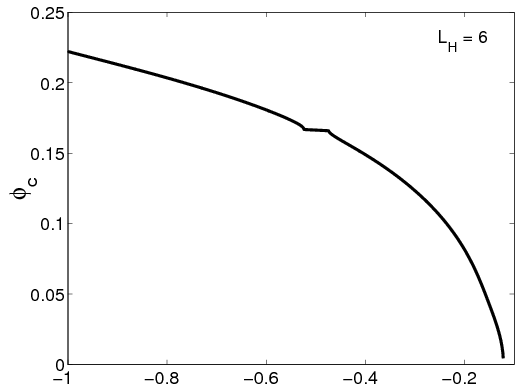}} 
\put(180,0){\includegraphics[width=6cm]{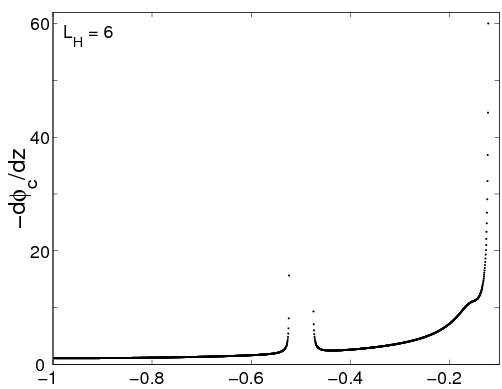}} 
\end{picture}
}
\end{center}
\caption{The normalized phase  $\phi_C(z)$ of the equimodular curve of
  $T_C(z;L_h)$ and the derivative $-d\phi_C(z)/dz$ for
  $L_h=6$ which has one gap on $-1\leq z \leq z_d$ where $\lambda_{\rm
  max}$ is real.}
\label{fig:philh6}
\end{figure}

The location of gaps for larger values of $L_h$ may be extrapolated  
by observing  that when the
maximum eigenvalues $\lambda_{\rm max}$ are complex they may be written 
as $|\lambda_{\rm max}|e^{\pm i\theta/2}$ where $\theta$ is defined in
section \ref{zerostophase}. The eigenvalues
collide and become real when $\theta/\pi$ is an integer. 
In principle each of the separate equimodular curves on 
$-1\leq z\leq z_d$ could be independent of each
other, but as long as we are to the right of any equimodular curve
which intersects the $z$ axis, we define by convention 
the eigenvalue phase at the right of a gap to be the same as the phase
at the left of the gap.  We then choose $\theta$ not to be restricted to
the interval $0$ to $\pi$ but to continuously increase as $z$ decreases
from $z_d$ to the first crossing of an equimodular curve. This
convention preserves the alternation of the signs of the real eigenvalues 
seen in table \ref{tab:gaps1}. For $L_h=6$ we illustrate the behavior of
this phase in figure \ref{fig:philh6}.
At the boundaries of the gaps the derivative
of the phase diverges as a square root, and for $L_h=6$ this derivative is
also plotted in figure \ref{fig:philh6}.

\begin{table}[!t]
\begin{center}
\begin{tabular}{|l||l|l|l|l|l|l|l|l|}\hline
$L_h$&1&2&3&4&5&6&7&8\\\hline 
$22$&$-0.152$&$-0.218$&$-0.346$&$-0.598$&&&&\\
$24$&$-0.148$&$-0.204$&$-0.305$&$-0.494$&$-0.844$&&&\\
$26$&$-0.145$&$-0.193$&$-0.276$&$-0.423$&$-0.683$&&&\\
$28$&$-0.143$&$-0.184$&$-0.254$&$-0.371$&$-0.574$&$-0.93$&&\\
$30$&$-0.140$&$-0.178$&$-0.237$&$-0.334$&$-0.495$&$-0.75$&&\\
\hline
$32$&$-0.1388$&$-0.172$&$-0.223$&$-0.305$&$-0.435$&$-0.642$&$-0.972$&\\
$34$&$-0.1373$&$-0.167$&$-0.213$&$-0.282$&$-0.390$&$-0.558$&$-0.815$&\\
$36$&$-0.1360$&$-0.163$&$-0.204$&$-0.264$&$-0.355$&$-0.494$&$-0.701$&\\
$38$&$-0.1348$&$-0.160$&$-0.196$&$-0.249$&$-0.327$&$-0.444$&$-0.616$&$-0.871$\\
$40$&$-0.1338$&$-0.157$&$-0.190$&$-0.237$&$-0.305$&$-0.405$&$-0.548$&$-0.752$\\
\hline
\end{tabular}
\end{center}
\caption{The location of the very small gaps on the segments
$-1\leq z \leq z_d$ where the maximum eigenvalue of the transfer
  matrix $T_C(z;L_h)$  for hard squares is real. For 
$L_h=22,~24,~26,~ 28,~30$ the values are obtained from the data; for
  $L_h\geq 32$ the values are obtained from extrapolation using figure
  \ref{fig:phi}. }
\label{tab:gaps2}
\end{table}

\begin{table}[!h]
\begin{center}
\begin{tabular}{|r|l|l|l|l|}\hline
$L_h$& $z_l(L_h)$&$z_r(L_h)$&gap&eigenvalue sign\\ \hline
$6$&$-0.4517$&$-0.4439$&$7.8\times 10^{-3}$&$-$\\
$8$&$-0.3004$&$-0.2999$&$5\times 10^{-4}$&$-$\\
$10$&$-0.23987$&$-0.23983$&$4\times 10^{-5}$&$-$\\
&$-0.6933$&$-0.6868$&$6.6\times 10^{-3}$&$+$\\
$12$&$-0.2079551$&$-0.2079504$&$4.6\times 19^{-6}$&$-$\\
&$-0.46977$&$-0.46908$&$6.9\times 10^{-4}$&$+$\\
$14$&$-0.18864888$&$-0.8864835$&$5.3\times 10^{-7}$&$-$\\
&$-0.362749$&$-0.362722$&$2.7\times 10^{-5}$&$+$\\
&$-0.85376$&$-0.85315$&$6.1\times 10^{-4}$&$-$\\
$16$&$-0.175819604$&$-0.175819540$&$6.4\times 10^{-8}$&$-$\\
&$-0.3024077$&$-0.3024052$&$2.5\times 10^{-6}$&$+$\\
&$-0.61069$&$-0.61049$&$2.0\times 10^{-4}$&$-$\\
\hline
\end{tabular}
\end{center}
\caption{The gaps on the segment $-1\leq z \leq z_d$ where the maximum
  eigenvalue of the transfer matrix $T_F(z;L_h)$  for hard squares on
  the free chain  of length $L_h$ is real for $6\leq L_h \leq 16$.}
\label{tab:gaps1f}
\end{table}

\begin{figure}[!h]
\begin{center}
\hspace{0cm} \mbox{
\begin{picture}(350,150 )
\put(0,0){\includegraphics[width=6cm]{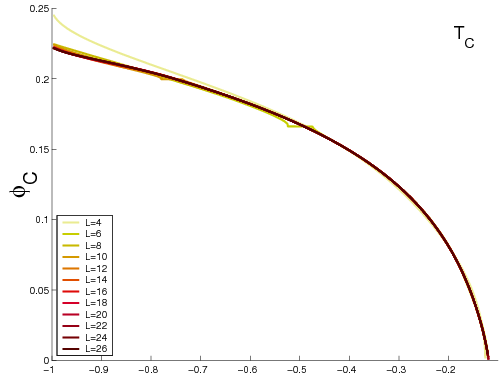}} 
\put(180,0){\includegraphics[width=6cm]{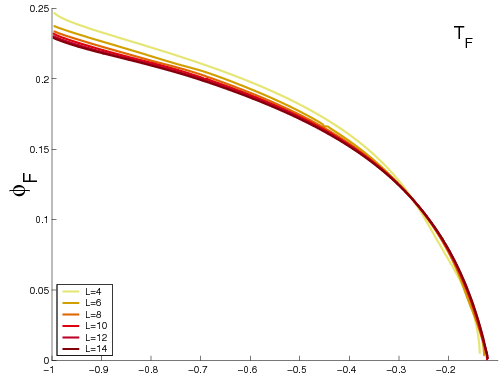}} 
\end{picture}
}
\end{center}
\caption{The normalized phase angles $\phi_C$ of $T_C(z;L_h)$ (on the
  left) and $\phi_F$ of $T_F(z;L_h)$ (on the right) on the
  segment $-1\leq z \leq z_d$ as a function of $z$.}
\label{fig:phi}
\end{figure}

For any given value of $z$ this unrestricted phase grows linearly with $L_h$
and thus we define a normalized phase
\begin{equation}
\phi=\frac{\theta}{2\pi L_h}.
\label{phasephi}
\end{equation}
The gaps occur when $L_h\phi=1$. 
In figure \ref{fig:phi} we plot the normalized phases $\phi_C$ of
$T_C(z;L_h)$
for  $4\leq L_h\leq 26$ and observe that they fall remarkably 
close to a common limiting
curve. We may thus use this curve to extrapolate the locations of the
gaps for  $L_h\geq 32$. These values are given in table
\ref{tab:gaps2} for $32\leq L_h \leq 40$.  
We also plot in figure \ref{fig:phi} the normalized phase $\phi_F$ for
$T_F(z;L_h)$ and note that $\phi_F\rightarrow \phi_C$ as $L_h$ becomes large.

\subsection{The density of partition zeros  of $L\times L$ lattices 
on the negative $z$ axis}
\label{density}

For both hard squares and
hard hexagons the zeros on the negative real axis are sufficiently
dense that a  quantitative comparison in terms of a density is possible.

The density of partition function zeros  on $L_v\times L_h$ lattices 
with $L_v/L_h$ fixed and $L_v,~L_h\rightarrow \infty$ 
is the limit 
of the finite lattice quantity
\begin{equation}
{\tilde D}_{L_v,L_h}(z_j)=\frac{1}{L_vL_h(z_{j+1}-z_j)}>0
\label{densdef}
\end{equation}
and the positions of the zeros $z_j$ increase monotonically with
$j$. To analyze this density we will  also need the $n^{\rm th}$ order lattice
derivative
\begin{equation} 
{\tilde D}^{(n)}_{L_v,L_h}(z_j)=\frac{{\tilde D}^{(n-1)}_{L_v,L_h}(z_{j+1})
-{\tilde D}^{(n-1)}_{L_v,L_h}(z_j)}{z_{j+1}-z_j}.
\end{equation}

As long as the 
density on $-1\leq z \leq z_d$  is the boundary of the zero free
region which includes the positive real axis (and where the thermodynamic 
limiting free energy is independent of the aspect ratio $L_v/L_h$), the
limiting density computed directly for the $L_v\times L_h$ lattice is
given in terms of the normalized phase angle (\ref{phasephi}) $\phi(z)$ 
on the interval $-1\leq z \leq z_d$ by use of (\ref{phidensity}) as
\begin{equation}
\lim_{L_h,L_v\rightarrow \infty}{\tilde
  D}_{L_v,L_h}(z)=-\lim_{L_h\rightarrow \infty}\frac{d \phi(z)}{dz}.
 \label{pzequality}
\end{equation}

\begin{figure}[!h]

\begin{center}
\hspace{0cm} \mbox{
\begin{picture}(380,300) 
\put(0,150){\includegraphics[width=6cm]{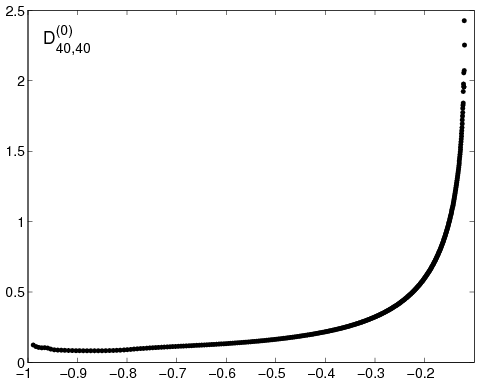}}
\put(190,150){\includegraphics[width=6cm]{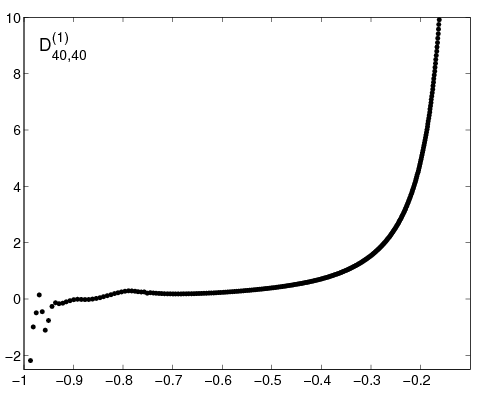}} 
\put(0,0){\includegraphics[width=6cm]{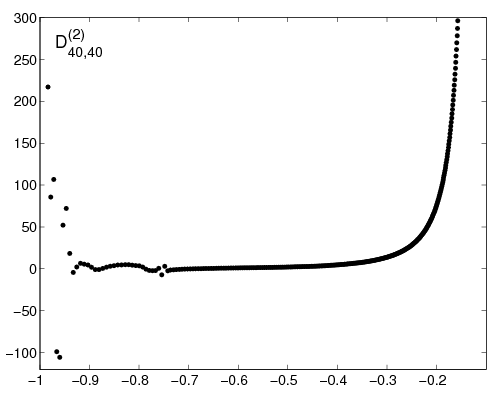}}
\put(190,0){\includegraphics[width=6cm]{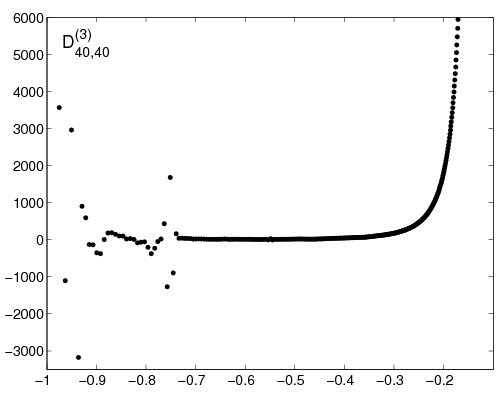}}
\end{picture}
}
\end{center}
\caption{The density of  zeros and the first three lattice derivatives 
for hard squares for the $40\times 40$ lattice with cylindrical
boundary conditions in the region $-1\leq z \leq z_d$. The glitch,
defined in section \ref{glitch},
caused by the gap given in table \ref{tab:gaps2} at $z=-0.752$ is
clearly visible in the second and third derivatives. } 
\label{fig:der40}

\end{figure}

Partition function zeros have been computed for systems much larger
than it has been possible to  compute eigenvalues  and the
largest lattices are for  the $L\times L$ cylinders. 
In figure 
\ref{fig:der40} we plot the density and the first three lattice
derivatives for hard squares for the $40\times 40$ cylindrical 
lattice on $-1\leq z \leq z_d$. On this scale the density appears to 
be quite smooth 
and a local maximum is seen in the first derivative.

\subsection{Partition zeros versus phase derivatives}

For hard hexagons the density of partition function zeros on the
negative $z$ axis lie very
close to the density computed from the derivative of the phase angle 
(\ref{pzequality}). Moreover  all the lattice derivatives are smooth
and featureless except very near $z_{d;hh}$ and also agree remarkably
well with the derivatives computed from the phase angle. This is in
significant contrast to hard squares.

\begin{figure}[!h]

\begin{center}
\hspace{0cm} \mbox{
\begin{picture}(400,300)
\put(0,0){\includegraphics[width=12cm]{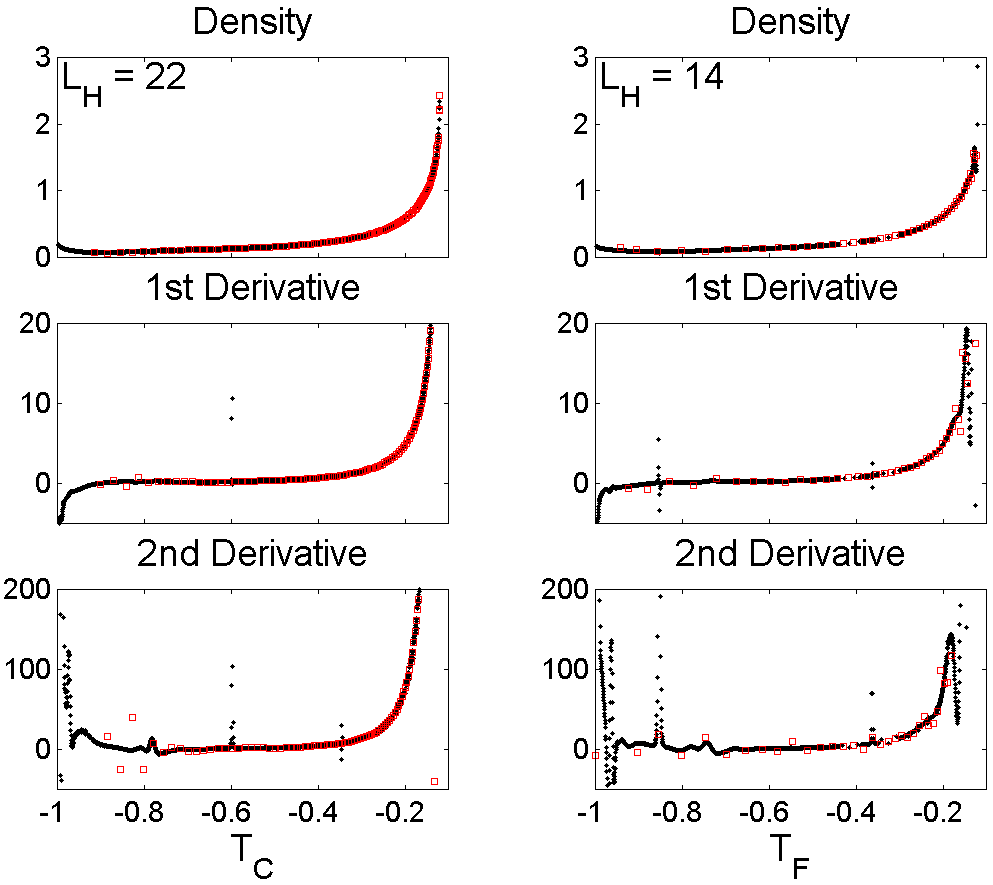}}
\end{picture}
}
\end{center}
\caption{The density and the  first two derivatives 
of the partition function zeros (in red) compared with the derivatives
of the normalized phase derivative curves (in black) of the toroidal 
lattice $Z^{CC}_{22,22}(z)$ for the $T_C(z;22)$ on the left and  the zeros of 
$Z^{CF}_{14,14}(z)=Z^{FC}_{14,14}(z)$ cylinder and
the  $T_F(z;14)$ transfer matrices (on the right). 
The divergences due to the gaps at 
$z=-0.598,~-0.346$ for $T_C(z;22)$ and at $z=-0.853,~-0.3627$ 
for $T_F(z;14)$ can be seen.}
\label{fig:dphcomp20}
\end{figure}

In figure \ref{fig:dphcomp20} 
we compare the density of zeros and its first two lattice derivatives
with the same quantities computed from the normalized  phase
derivative curves of
the corresponding transfer matrix for the $22\times 22$ toroidal
lattice and the $14\times 14$ cylindrical lattice.  For the density 
almost all zeros are seen to
fall remarkably close to the normalized phase derivative curves. In the first
derivative of the normalized phase derivative curve we see the divergences due 
to the gaps at $-0.60$ for $T_C(z;22)$
and at $-0.85$ and $-0.36$ for $T_F(z;14)$. In the second derivative,
the divergences become more pronounced and the gap at $-0.35$ 
of $T_C(z;22)$ becomes noticeable.

The derivatives of the normalized phase derivative 
curves all exhibit oscillations 
in the vicinity of $z=-1$ which
become larger and cover an increasing segment of the $z$ axis as
the order of the derivative increases. In these oscillatory regions
noticeable  discrepancies between the lattice derivative of the zeros and
the derivatives of the normalized phase are apparent.

\subsection{Glitches in the density of zeros}
\label{glitch}

The gaps in the equimodular curves of hard squares on $-1\leq z
\leq z_d$ which caused the divergences in the normalized phase curves in
figure  \ref{fig:dphcomp20}  lead to irregularities in the density 
of the $L\times L$ partition function zeros which we refer to as ``glitches''. 
These glitches upset the smoothness of the density of zeros on the 
finite lattice and  become increasingly apparent in the higher 
derivatives of the  density.   
The glitch at $z=-0.752$ is quite visible in the second and third
derivatives in figure \ref{fig:der40}.

\begin{figure}[!h]

\begin{center}
\hspace{0cm} \mbox{
\begin{picture}(300,230)
\put(0,125){\includegraphics[width=5cm]{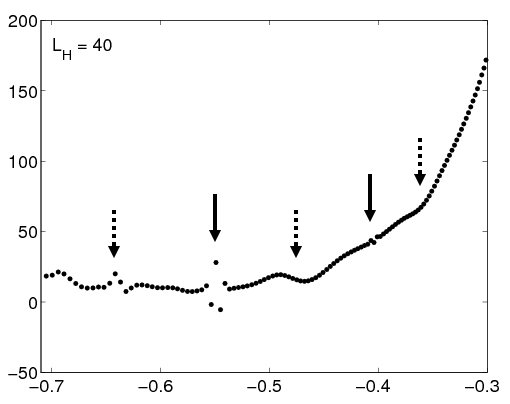}}
\put(150,125){\includegraphics[width=5cm]{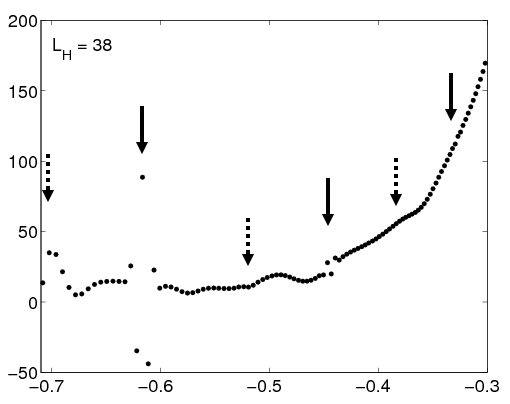}}
\put(0,0){\includegraphics[width=5cm]{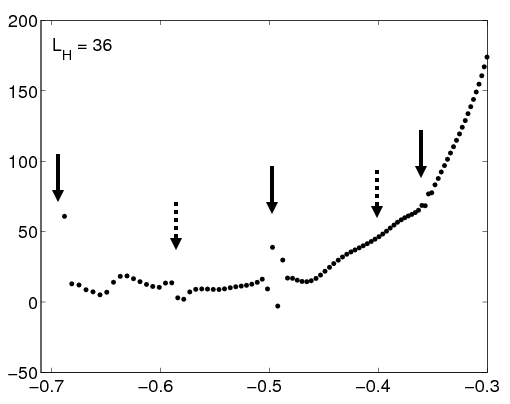}}
\put(150,0){\includegraphics[width=5cm]{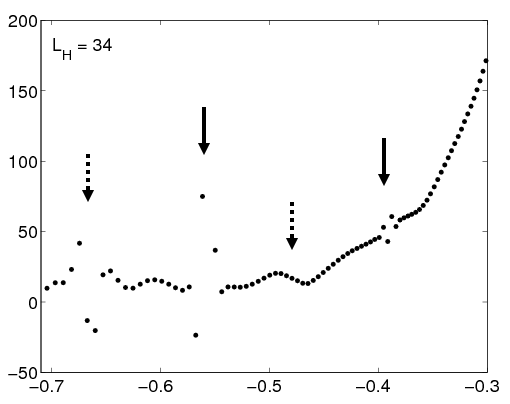}}
\end{picture}
}
\end{center}
\caption{The third derivative  of the density of  hard squares for
  $40\times 40,~38\times 38,~ 36\times 36,~34\times 34$ lattices with
  cylindrical boundary conditions 
 in the region 
$-0.7 \leq z \leq -0.25$ . The gaps of table \ref{tab:gaps2} are
  indicated by solid arrows} 
\label{fig:expder3}

\end{figure}

To illustrate further the relation of gaps to  glitches in the density of
zeros we plot the 
third derivatives of the density of cylindrical $L\times L$ lattices
on an expanded scale in figure \ref{fig:expder3} where we indicate   
with solid arrows the positions of the corresponding gaps in 
the $T_C(z;L_h)$ equimodular curves of table \ref{tab:gaps2}.
On these expanded scales we observe that as the size of the $L\times L$ lattice
increases the number of glitches increases, they move to the right and
their amplitude decreases. These properties follow from the properties
of the gaps of table \ref{tab:gaps2} and the normalized phase curve of 
figure \ref{fig:phi}.

There also appear to be deviations of the zeros from a smooth curve at
values of $z$ where the phases of the complex conjugate pair of
maximum modulus eigenvalues are $\pm \pi/2$. These deviations have no
relation to gaps in the equimodular curves and are indicate with
dashed arrows in figure \ref{fig:expder3}.

\subsection{Hard square density of zeros for  $z\rightarrow z_d$.}
\label{sec:hszd}

\begin{figure}[!h]

\begin{center}
\hspace{0cm} \mbox{
\begin{picture}(150,150)
\put(0,0){\includegraphics[width=7cm]{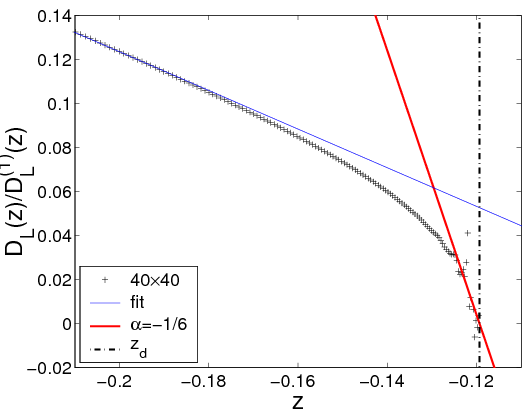}}
\end{picture}
}
\end{center}
\caption{The plot of density/derivative for the partition function zeros 
of hard squares for $40\times
  40$ cylindrical lattice. The red line
has $\alpha=1/6$ and $z_d=-0.119$. The blue line has 
$\alpha_f=1/0.88=1.14$ and $z_f=-0.058$.}
\label{fig:nearzd}
\end{figure}

As $z\rightarrow z_d$ the density diverges as
\begin{equation}
D(z)\sim (z_d-z)^{-\alpha},
\end{equation}
where from the universality of the point $z_d$ with the Lee-Yang edge
it is expected that $\alpha=1/6$, which was also found to be the case for
hard hexagons. We investigate the exponent
$\alpha$ using the method used in \cite{assis}  
by plotting in figure \ref{fig:nearzd} the quantity
$D_L(z)/D^{(1)}_L(z)$ for $L=40$ and compare this with
\begin{equation}
D(z)/D'(z)\sim (z_d-z)/\alpha~~{\rm with}~~ \alpha=1/6,
\end{equation}
which is expected to hold for $z\rightarrow z_d$.

As was the case for hard hexagons this limiting form 
is seen to hold only for $z$ very close to
$z_d$ and for comparison we also plot a fitting function
\begin{equation}
f(z)=(z_f-z)/\alpha_f ~~{\rm with}~~
z_f=-0.058,~~\alpha_f=1/0.88,
\end{equation}
which well approximates the curve in the range $-0.30\leq z \leq
-0.16$. This same phenomenon has been seen in  \cite[equation
    (4.8) and figure 41]{jesper} for Hamiltonian chains.

\subsection{The point $z=-1$} 

Hard squares have the remarkable property, which has no counterpart
for hard hexagons, that
at $z=-1$ all roots of the characteristic equation are either roots of
one, or minus one, with various multiplicities. These roots have been computed 
for the full transfer matrix $T_C(-1;L_h)$ either directly \cite{fse}, 
\cite{baxneg}
to size $15$ or using a mapping to rhombus tilings \cite{jon1} 
to size $L_h=50$. In  \ref{appa}  we present  factorizations of 
the characteristic polynomial $T_C(-1;L_h)$ for the reduced sector 
$P=0^+$ for $L_h\leq 29$, and of $T_F(-1;L_h)$ for $L_h\leq 20$ 
both for the unrestricted and positive parity sectors.
In \ref{appb} we give the
partition function values at $z=-1$.

\subsection{Behavior near $z=-1$}

The density of zeros of figure \ref{fig:der40} for the $40\times 40$
 cylinder is finite as
$z\rightarrow -1$. However  the first derivative is sufficiently
scattered for $z\leq -0.95$ that an estimate of the slope is impracticable.

\begin{figure}[!h]

\begin{center}
\hspace{0cm} \mbox{
\begin{picture}(290,180)
\put(0,30){\includegraphics[width=6cm]{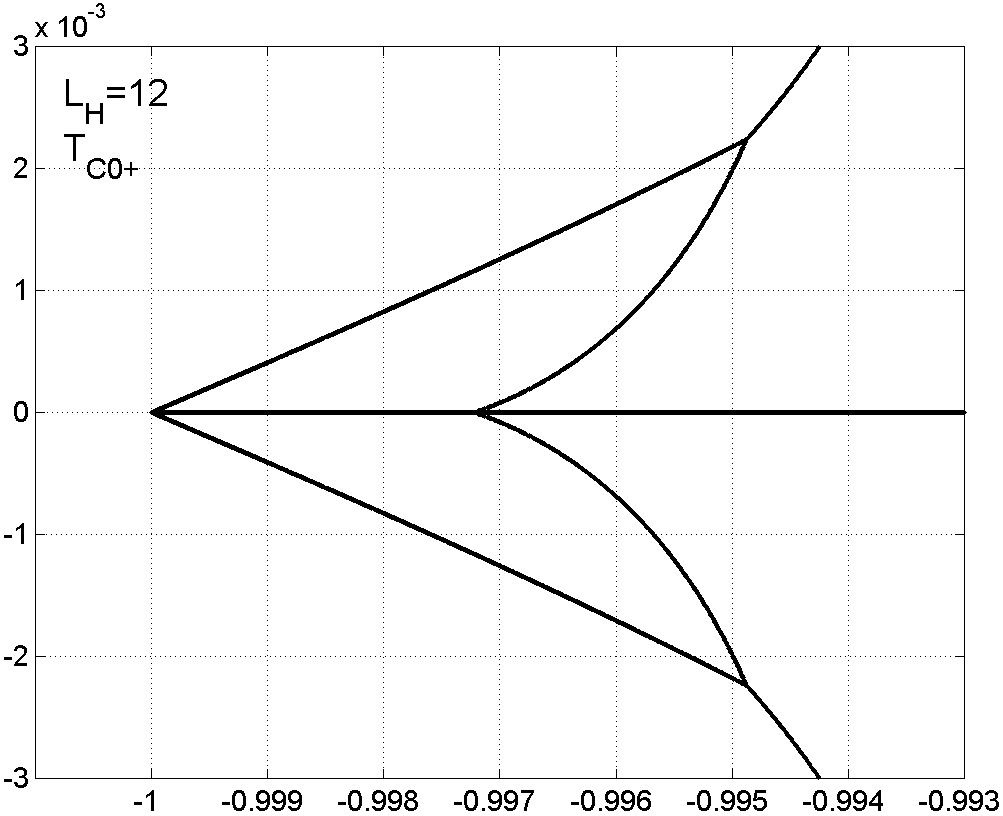}}
\put(190,0){\includegraphics[width=4cm]{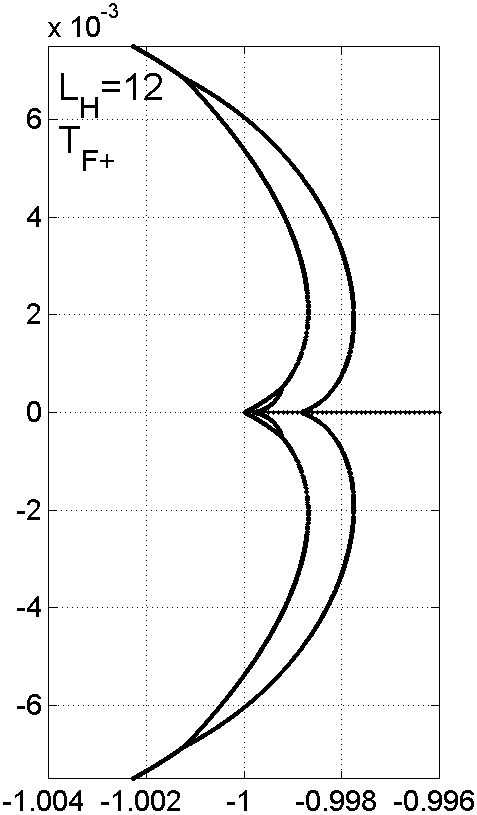}}
\end{picture}
}
\end{center}
\caption{Plots in the complex fugacity $z$ plane near $z=-1$  for
  $L_h=12$ of the equimodular curves of hard square transfer matrix
  $T_C(z;L_h)$ with $P=0^+$ (on the left) and $T_F(z,L_h)$ with $+$
  parity (on the right) on a
  scale which shows the level crossings on the $z$-axis to the right
  of $z=-1$.}
\label{fig:hszm112}

\end{figure}

Furthermore there is a great amount of structure in the equimodular curves
near the point $z=-1$ where all eigenvalues are 
equimodular and which is not apparent on the scale of the plots in figure
\ref{fig:evzeroshs}. We illustrate this 
complexity for $L_h=12$ for
$P=0^+$ in figure \ref{fig:hszm112} where we see that there are
equimodular curves which intersect the $z$ axis for $z\geq -1$. These
level crossings are a feature also for $T_C(z)$ without the
restriction to $P=0^+$ and for $T_F(z)$ and $T_F(z)$ with $+$ parity as
well. In general there are
several  such crossings for a given $L_h$. We give the values of
the crossing furthest to the right  in table \ref{tab:crossing}.
It is not clear whether these level crossings will persist to the
right of $z=-1$ as $L_h\rightarrow\infty$. We also note that often
there are more than one such level crossing, as illustrated in
figure~\ref{fig:hszm112} for $T_F(z;12)$.

\begin{table}[!h]
\begin{center}
\begin{tabular}{|r|l|l|l|l|l|}\hline
$L$& $P=0^+$&$T_C$&&parity $=+$&$T_F$\\ \hline
$12$&$-0.9973$&$-0.91295$&&$-0.9988$&same\\
$14$&none&$-0.9195$&&$-0.999296$&$-0.999092$\\
$16$&none&$-0.96$&&none&none\\
$18$&$-0.99994$&&&$-0.9990$&\\
$20$&$-0.9995$&&&&\\
$22$&$-0.9999$&&&&\\
$24$&$-0.9974$&&&&\\
$26$&$-0.9990$&&&&\\
$28$&$-0.9996$&&&&\\
\hline
\end{tabular}
\end{center}
\caption{Positions of the right-most  equimodular curve crossings of 
the negative $z$-axis for hard squares of $T_C(z)$ in 
the sector $P=0^+$ and unrestricted and of $T_F(z)$ in the plus parity
sector and unrestricted. } 
\label{tab:crossing}
\end{table}

\section{Discussion}
\label{discussion}

The three different techniques of series expansions, transfer matrix
eigenvalues and partition function zeros give three quite different
perspectives on the difference between the integrable hard hexagon
model and non-integrable hard squares. 

\subsection{Series expansions}

Consider first the series expansion of the
physical free energy of hard squares \cite{chan,jensen}, which is analyzed by means of differential approximants, 
as compared with the exact solution of hard hexagons \cite{baxterhh}.

The hard hexagon free energy for both the high and low density regimes
satisfies  Fuchsian differential
equations  which can be obtained from a finite number of
terms in a  series expansion \cite{assis}. 

For an non-integrable model like hard squares, the best kind of
differential approximant analysis to be introduced is not clear. For
integrable models,
even if one has a small number of series coefficients, restricting to
Fuchsian ODEs
has been seen to be an extremely efficient constraint. However for a
(probably non-integrable)
model like hard squares, there is no reason to restrict the
linear differential equations
annihilating the hard square series to be Fuchsian.
In \cite{jensen} the existing 92 term series are
analyzed by means of differential approximants but the series is too 
short to determine whether $z=-1$ is, or is not, a singular point.



The method of series expansions and differential approximants are not
well adapted to analyze qualitative differences between hard squares
and hard hexagons. This is to be compared with the transfer matrix
eigenvalues and partition function zeros presented above which show
dramatic differences between the two systems.

\subsection{Transfer matrices}

The clearest distinction between integrable hard hexagons and
non-integrable hard squares is seen in the factorization properties of
the discriminant of the characteristic polynomials of the transfer matrices
$T_C(z;L_h)$ and $T_F(z;L_h)$. At the zeros of the discriminant the
transfer matrix in general  fails to be diagonalizable and the eigenvalues may
have singularities. 

For hard hexagons these discriminants
contain square factors which exclude the existence of gaps in the
equimodular curves and singularities of the maximum eigenvalue on 
the negative $z$-axis. This was observed for
$T_C(z;L_h)$ in \cite{assis}. In the  present paper 
 these square factors
 and lack of gaps has been observed for the transfer matrix
$T_F(z;L_h)$ of hard hexagons for all value of $L_h$
 studied and supports the conjecture that integrability
 can be established by extending the methods of \cite{cher}-\cite{ahn}.
For hard squares there are no such factorizations, so that its
equimodular curves have gaps and the maximum eigenvalue has
singularities on the negative real $z$-axis.

\subsection{Partition function zeros}

In  \cite{assis} we qualitatively characterized the partition function
zeros as either being on curves or being part of a necklace, and in
the present paper we have characterized the zeros as filling up areas.
However, further investigation is required to determine if these 
characterizations of the qualitative appearance of zeros of the finite system 
characterize the thermodynamic limit. In \cite{assis} we initiated
such a study by examining the dependence of the right-hand endpoints
of the necklace on the size of the lattice and observed that the
endpoints move to the right as the lattice size increases. However,
there is not sufficient data to reliably determine the limiting
behavior. Thus, if in the thermodynamic limit the endpoint moved to
$z_{c;hh}$ the notion of zeros being on a curve might not persist.
Similarly, it needs further investigation to determine if the zeros 
of hard squares, which we have characterized as filling up an area, 
will  fill the area in the thermodynamic limit or whether
further structure develops.

On the negative $z$-axis  both hard hexagons and hard squares
have a line of zeros which has been investigated in detail in section 
\ref{negz}. The density of zeros for $z<z_{d:hh}$ for  hard
hexagons  is mostly featureless and smooth,
which is quite consistent with the low density free energy having a
branch cut starting at $z_{d;hh}$. Hard squares zeros, on the other
hand, have a series of ``glitches'' whose number increases as $z$
approaches $z_d$ and which correspond to the locations of the gaps in
the equimodular curves. A rigorous analysis of behavior of these glitches
needs to be made.

\subsection{Behavior near $z_c$}

The equimodular curves of hard hexagons  were extensively studied in
\cite{assis}. The equimodular curves, as illustrated for
$L_h=21$ in figure  \ref{fig:evzeroshh}, 
consist of the curve where the low and high
density physical free energy are equimodular and a necklace region
which surrounds this equimodular curve in part of the left half-plane.

\begin{figure}[!h]

\begin{center}
\hspace{0cm} \mbox{
\begin{picture}(250,250)
\put(0,0){\includegraphics[width=8cm]{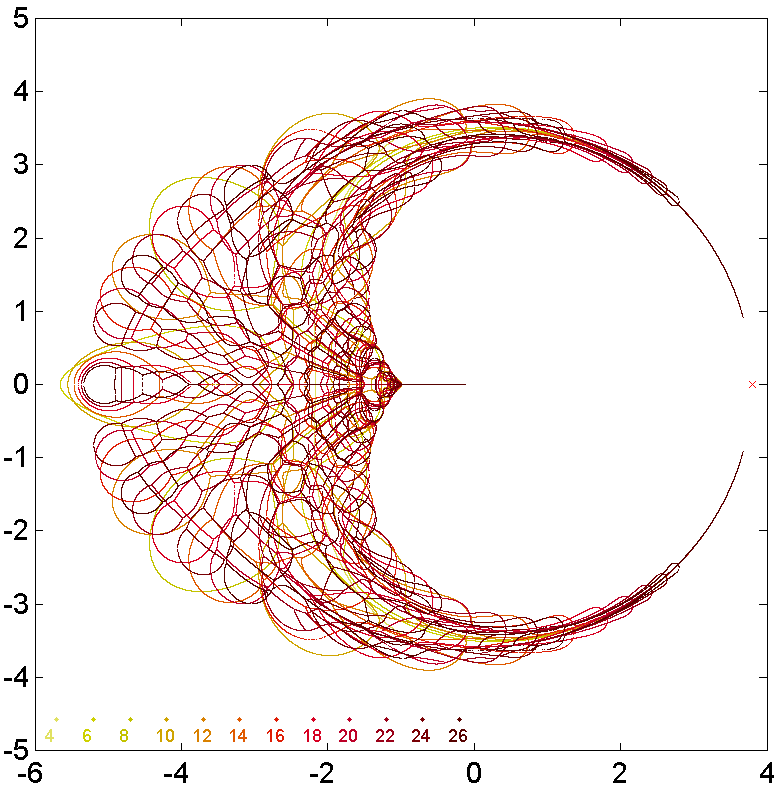}}
\end{picture}
}
\end{center}
\caption{The equimodular curves in the complex $z$ plane of the 
$T_C(z;L_h)$ transfer matrix in
  the $0^+$ sector for $4\leq L_h\leq 26$ plotted together. The
  different values of $L_h$ are given different shadings as indicated
  on the plot. The location of $z_c$ is indicated by a cross.}
\label{fig:equimodall}

\end{figure}

For hard hexagons there is only one unique curve of zeros of the
$L\times L$  partition function which is converging towards 
$z_{c;hh}$ as $L\rightarrow \infty$. However, for hard squares the  
partition function
zeros in  figures \ref{fig:cf}-\ref{fig:free} do not lie on a
single unique curve near $z_c$. This is clearly seen in the plots of
figure \ref{fig:hsall} where the
zeros appear to be converging to a wedge behavior as $L\rightarrow
\infty$ which is analogous to the behavior of the equimodular curves
of figure \ref{fig:equimodall}.

The behavior of the equimodular curves of hard squares near $z_c$
in figure \ref{fig:evzeroshs} is qualitatively different from the
behavior of hard hexagons in figure \ref{fig:evzeroshh}. This is
vividly illustrated in figure \ref{fig:equimodall} where we plot the
equimodular curves for $T_c(z;L_h)$ with $P=0^+$ for all values $4\leq
L_h \leq 26$. We see in this figure that there is an ever increasing set of
loops in the equimodular curves which approach $z_c$ as
$L_h\rightarrow \infty$.

It needs to be investigated if this behavior of both the
zeros and the equimodular curves for hard squares will have an effect on the
singularity at $z_c$ beyond what is obtained from the analysis of
the series expansion of \cite{chan,jensen}.

\subsection{Behavior near $z=-1$}

Finally we note that the relation of the equimodularity of all
eigenvalues at $z=-1$ to the analytic behavior of the physical free energy
is completely unknown, as is the curious observation  for 
$12\leq L_h\leq 28$  found in table \ref{tab:crossing} that there are
equimodular curves of $T_C(z;L_h)$ and for $T_F(z;L_h)$ 
which cross the negative $z$-axis to the right of
$z=-1$. There are many values of $L_h$ for which there are more than
one such curve. It would be of interest to know if this feature persists for
$L_h>28$ and if it does, does the point of  rightmost 
crossing move to the right.
If such a phenomenon does exist it would cause a re-evaluation of
the role of zeros on the negative $z$-axis.

\section{Conclusions}
\label{conclusions}
The techniques of series expansions, universality and the 
renormalization group apply equally well to describe the
dominant behavior at $z_c$ and $z_d$ of hard hexagons and hard squares. 
However the results of this paper reveal many differences between integrable
hard hexagons and non-integrable hard squares which have the potential 
to create further analytic properties in hard squares which are not
present in hard hexagons.

The renormalization group combined with conformal field theory
predicts that both $z_c$ and $z_d$ will be isolated regular singularities
where the free energy will have a finite number of algebraic or
logarithmic singularities, each multiplied by a convergent infinite
series. This scenario is, of course, far beyond what can be
confirmed by numerical methods. Indeed  hard squares are predicted to
have the same  set of 5 exponents at $z_d$ which hard hexagons have
\cite{joyce},\cite{assis}
even though only two such exponents can be obtained from the 92 terms
series expansion \cite{jensen}.  

The emergence of the critical singularities predicted by the
renormalization group at either $z_c$ or $z_d$ is a
phenomenon which relies upon the thermodynamic
limit and we have seen that hard squares approach this limit in a 
more complicated manner than do hard hexagons.

Near $z_c$ the limiting position of the zeros for hard squares 
appears to be a wedge. 
This is far more complex than the behavior of hard hexagons.

Near $z_d$ the zeros of both hard hexagons and hard squares are
observed to lie on a segment of the negative $z$-axis. If this 
indeed holds in the
thermodynamic limit it would be satisfying if a genuine proof could be
found which incorporates the fact that some level crossings have been
observed to the right of $z=-1$.

On the negative $z$-axis hard squares have glitches in the density of
zeros and gaps in the equimodular curves which hard hexagons do 
not have. In the thermodynamic limit the
glitches and gaps may become a dense set of measure zero by the analysis
leading to table \ref{tab:gaps2} . Does this give a hint of 
the analytical structure of non-integrable models?


\vspace{.3cm}

{\bf  Acknowledgments} 

\vspace{.1in}

We are pleased to acknowledge fruitful discussions 
with C. Ahn, A.J. Guttmann, and P.A. Pearce. One of us (JJ) is pleased to thank
the Institut Universitaire de France and Agence Nationale de la
Recherche under grant ANR-10-BLAN-0401 and the Simons Center for
Geometry and Physics for their hospitality. One of us (IJ) was
supported by an award under the Merit Allocation Scheme of the NCI
National facility at the ANU  and by funding under the
Australian Research Council's Discovery Projects scheme by the grant
DP140101110. We also made extensive use of the High Performance
Computing services offered by ITS Research Services at the University
of Melbourne.

\appendix

 \section{Characteristic polynomials at $z=-1$}
\label{appa}

In~\cite{jon1} it was proven that all of the eigenvalues of the
$T_C(-1;L_h)$ transfer matrix at $z=-1$ are roots of unity and the
characteristic polynomials were given in that paper up to $L_h=50$. 
Below we give the
factorized characteristic polynomials $P^{C0+}_{L_h}$ in the $0^+$
sector at $z=-1$ up to $L_h=29$. The transfer matrix $T_F(z;L_h)$ has
not been considered before in the literature, and below we give the
factorized characteristic polynomials $P^F_{L_h}$ and $P^{F+}_{L_h}$ of
the full $T_F(-1;L_h)$ and the restricted positive parity sector,
respectively, at $z=-1$ up to $L_h=20$. In all cases divisions are
exact.

\subsection{Characteristic polynomials $P_{L_h}^F$}
 
The degree of $P_{L_h}^F$ is exactly the 
Fibonacci number $F(n)$ defined by the
recursion relation
\begin{equation}
F(L_h+2) = F(L_h+1) + F(L_h)
\label{fib}
\end{equation}
with the initial conditions $F(-1)=0$, $F(0)=1$, so that its generating function is
\begin{equation}
G^{F} = \frac{(2+t)}{(1-t-t^2)}
\end{equation}
and thus as $L_h\rightarrow \infty$ the degree of the polynomial $P_{L_h}^F$ 
grows as $N_G^{L_h}$, where $N_G=(1+{\sqrt  5})/2\sim 1.618\cdots$ is the golden ratio.

The first 20 polynomials are
\begin{eqnarray}
\hspace{-1.in} P^{F}_{1} &=&
(x^{6}-1)(x^{3}-1)^{-1}(x^{2}-1)^{-1}(x-1)
\nonumber\\
\hspace{-1.in} P^{F}_{2} &=& (x^{4}-1)(x^{2}-1)^{-1}(x-1)\nonumber\\
\hspace{-1.in} P^{F}_{3} &=& (x^{8}-1)(x^{4}-1)^{-1}(x-1)\nonumber\\
\hspace{-1.in} P^{F}_{4} &=& (x^{6}-1)(x^{4}-1)(x^{3}-1)^{-1}(x-1)\nonumber\\
\hspace{-1.in} P^{F}_{5} &=& (x^{10}-1)(x^{8}-1)/(x^{4}-1)(x^{2}-1)^{-1}(x-1)\nonumber\\
\hspace{-1.in} P^{F}_{6} &=& (x^{14}-1)(x^{4}-1)^{2}(x^{2}-1)^{-1}(x-1)\nonumber\\
\hspace{-1.in} P^{F}_{7} &=& (x^{18}-1)(x^{12}-1)(x^{8}-1)(x^{6}-1)(x^{4}-1)^{-2}(x^{3}-1)^{-1}(x-1)\nonumber\\
\hspace{-1.in} P^{F}_{8} &=& (x^{22}-1)(x^{16}-1)^{2}(x^{8}-1)^{-1}(x^{4}-1)^{2}(x-1)\nonumber\\
\hspace{-1.in} P^{F}_{9} &=& (x^{26}-1)(x^{20}-1)^{3}(x^{14}-1)(x^{10}-1)^{-1}(x^{8}-1)(x^{4}-1)^{-2}(x^{2}-1)^{-1}(x-1)\nonumber\\
\hspace{-1.in} P^{F}_{10} &=& (x^{30}-1)(x^{24}-1)^{3}(x^{18}-1)^{2}(x^{8}-1)^{-1}(x^{6}-1)(x^{4}-1)^{3}(x^{3}-1)^{-1} \nonumber\\
\hspace{-1.in} &&\qquad (x^{2}-1)^{-1}(x-1)\nonumber\\
\hspace{-1.in} P^{F}_{11} &=& (x^{34}-1)(x^{28}-1)^{3}(x^{22}-1)^{4}(x^{16}-1)(x^{14}-1)(x^{8}-1)(x^{4}-1)^{-3}(x-1)\nonumber\\
\hspace{-1.in} P^{F}_{12} &=& (x^{38}-1)(x^{32}-1)^{4}(x^{26}-1)^{6}(x^{20}-1)^{2}(x^{10}-1)(x^{8}-1)^{-1}(x^{4}-1)^{3}(x-1)\nonumber\\
\hspace{-1.in} P^{F}_{13} &=& (x^{42}-1)(x^{36}-1)^{5}(x^{30}-1)^{8}(x^{24}-1)^{5}(x^{12}-1)(x^{10}-1)(x^{8}-1)^{2}(x^{6}-1) \nonumber\\
\hspace{-1.in} &&\qquad (x^{4}-1)^{-3}(x^{3}-1)^{-1}(x^{2}-1)^{-1}(x-1)\nonumber\\
\hspace{-1.in} P^{F}_{14} &=& (x^{46}-1)(x^{40}-1)^{5}(x^{34}-1)^{11}(x^{28}-1)^{11}(x^{22}-1)^{3}(x^{14}-1)^{-1}(x^{8}-1)^{-1} \nonumber\\
\hspace{-1.in} &&\qquad 
(x^{4}-1)^{4}(x^{2}-1)^{-1}(x-1)\nonumber\\
\hspace{-1.in} P^{F}_{15} &=& (x^{50}-1)(x^{44}-1)^{5}(x^{38}-1)^{14}(x^{32}-1)^{18}(x^{26}-1)^{8}(x^{22}-1)(x^{20}-1) \nonumber\\
\hspace{-1.in} &&\qquad (x^{16}-1)^{-2}(x^{8}-1)^{2}(x^{4}-1)^{-4}(x-1)\nonumber\\
\hspace{-1.in} P^{F}_{16} &=& (x^{54}-1)(x^{48}-1)^{6}(x^{42}-1)^{17}(x^{36}-1)^{25}(x^{30}-1)^{17}(x^{24}-1)^{4}(x^{18}-1)(x^{14}-1) \nonumber\\
\hspace{-1.in} &&\qquad (x^{12}-1)^{-1}(x^{10}-1)^{-1}(x^{8}-1)^{-1}(x^{6}-1)(x^{4}-1)^{4}(x^{3}-1)^{-1}(x-1)\nonumber\\
\hspace{-1.in} P^{F}_{17} &=& (x^{58}-1)(x^{52}-1)^{7}(x^{46}-1)^{21}(x^{40}-1)^{35}(x^{34}-1)^{31}(x^{28}-1)^{11}
(x^{26}-1)^{-1} \nonumber\\
\hspace{-1.in} &&\qquad (x^{22}-1)(x^{20}-1)^{3}(x^{14}-1)^{-1}(x^{10}-1)^{-1}(x^{8}-1)^{2}(x^{4}-1)^{-4}(x^{2}-1)^{-1} \nonumber\\
\hspace{-1.in} &&\qquad (x-1)\nonumber\\
\hspace{-1.in} P^{F}_{18} &=& (x^{62}-1)(x^{56}-1)^{7}(x^{50}-1)^{25}(x^{44}-1)^{50}(x^{38}-1)^{52}(x^{32}-1)^{24}(x^{26}-1)^{4} \nonumber\\
\hspace{-1.in} &&\qquad (x^{22}-1)^{-1}(x^{16}-1)^{2}(x^{8}-1)^{-2}(x^{4}-1)^{5}(x^{2}-1)^{-1}(x-1)\nonumber\\
\hspace{-1.in} P^{F}_{19} &=& (x^{66}-1)(x^{60}-1)^{7}(x^{54}-1)^{29}(x^{48}-1)^{67}(x^{42}-1)^{82}(x^{36}-1)^{50}(x^{30}-1)^{14} \nonumber\\
\hspace{-1.in} &&\qquad (x^{24}-1)^{-2}(x^{18}-1)^{3}(x^{14}-1)^{-1}(x^{12}-1)(x^{10}-1)(x^{8}-1)^{2}(x^{6}-1) \nonumber\\
\hspace{-1.in} &&\qquad 
(x^{4}-1)^{-5}(x^{3}-1)^{-1}(x-1)\nonumber\\
\hspace{-1.in} P^{F}_{20} &=& (x^{70}-1)(x^{64}-1)^{8}(x^{58}-1)^{34}(x^{52}-1)^{84}(x^{46}-1)^{122}(x^{40}-1)^{97}(x^{34}-1)^{35} \nonumber\\
\hspace{-1.in} &&\qquad (x^{28}-1)^{4}(x^{26}-1)(x^{20}-1)^{-3}(x^{14}-1)(x^{10}-1)(x^{8}-1)^{-2}(x^{4}-1)^{5} \nonumber\\
\hspace{-1.in} &&\qquad (x-1)
\end{eqnarray}
and from these we see that the degrees of the multiplicity of the
eigenvalue $+1$ are
\begin{equation}
\hspace{-1.in} 1, 1, 1, 2, 1, 3, 2, 5, 3, 8, 9, 17, 20, 33, 45, 74, 105, 167, 250,389,\ldots
\end{equation}
where we find  a ``mod 4'' effect.

\subsection{Characteristic polynomials $P_{L_h}^{F+}$}
The degrees of $P_{L_h}^{F+}$ follow the sequence A001224 in the 
OEIS~\cite{oeis}
and they are 
related to the Fibonacci sequence $F(n)$ as follows:
\begin{eqnarray}
\frac{F(L_h+1)+F(\frac{L_h+1}{2}+1)}{2}, &\qquad& L_h=\mathrm{odd}\\
\frac{F(L_h+1)+F(\frac{L_h}{2})}{2}, &\qquad& L_h=\mathrm{even}
\end{eqnarray}
This sequence has the following generating function
\begin{equation}
G^{F+}  = \frac{G^F}{2} + \frac{t^3+t^2+t+2}{2(1-t^2-t^4)}
\end{equation}
so that the degree of the polynomials $P_{L_h}^{F+}$ grow as $N_G^{L_h}$ with a sub-dominant growth of $N_G^{L_h/2}$.

The first 20 polynomials are
\begin{eqnarray}
\hspace{-1.in} P^{F+}_{1} &=& (x^{6}-1)(x^{3}-1)^{-1}(x^{2}-1)^{-1}(x-1)\nonumber\\
\hspace{-1.in} P^{F+}_{2} &=& (x^{4}-1)(x^{2}-1)^{-1}\nonumber\\
\hspace{-1.in} P^{F+}_{3} &=& (x^{8}-1)(x^{4}-1)^{-1}\nonumber\\
\hspace{-1.in} P^{F+}_{4} &=& (x^{6}-1)(x^{3}-1)^{-1}(x^{2}-1)\nonumber\\
\hspace{-1.in} P^{F+}_{5} &=& (x^{8}-1)(x^{5}-1)(x^{4}-1)^{-1}\nonumber\\
\hspace{-1.in} P^{F+}_{6} &=& (x^{7}-1)(x^{4}-1)^{2}(x^{2}-1)^{-2}(x-1)\nonumber\\
\hspace{-1.in} P^{F+}_{7} &=& (x^{18}-1)(x^{9}-1)^{-1}(x^{8}-1)(x^{6}-1)^{2}(x^{4}-1)^{-1}(x^{3}-1)^{-1}(x^{2}-1)^{-1} \nonumber\\
\hspace{-1.in} &&\qquad (x-1)\nonumber\\
\hspace{-1.in} P^{F+}_{8} &=& (x^{22}-1)(x^{16}-1)^{2}(x^{11}-1)^{-1}(x^{8}-1)^{-2}(x^{2}-1)(x-1)\nonumber\\
\hspace{-1.in} P^{F+}_{9} &=& (x^{20}-1)^{3}(x^{14}-1)(x^{13}-1)(x^{10}-1)^{-3}(x^{8}-1)(x^{7}-1)^{-1}(x^{4}-1)^{-2} \nonumber\\
\hspace{-1.in} &&\qquad (x-1)\nonumber\\
\hspace{-1.in} P^{F+}_{10} &=& (x^{18}-1)^{2}(x^{15}-1)(x^{12}-1)^{3}(x^{9}-1)^{-2}(x^{6}-1)(x^{4}-1)^{2}(x^{3}-1)^{-1} \nonumber\\
\hspace{-1.in} &&\qquad (x^{2}-1)^{-2}\nonumber\\
\hspace{-1.in} P^{F+}_{11} &=& (x^{34}-1)(x^{17}-1)^{-1}(x^{16}-1)(x^{14}-1)^{4}(x^{11}-1)^{4}(x^{4}-1)^{-1}(x^{2}-1)^{-1}\nonumber\\
\hspace{-1.in} P^{F+}_{12} &=& (x^{38}-1)(x^{32}-1)^{4}(x^{19}-1)^{-1}(x^{16}-1)^{-4}(x^{13}-1)^{6}(x^{10}-1)^{3}(x^{2}-1)^{2}\nonumber\\
\hspace{-1.in} P^{F+}_{13} &=& (x^{36}-1)^{5}(x^{30}-1)^{8}(x^{21}-1)(x^{18}-1)^{-5}(x^{15}-1)^{-8}(x^{12}-1)^{6}
(x^{10}-1)(x^{9}-1) \nonumber\\
\hspace{-1.in} &&\qquad (x^{8}-1)(x^{6}-1)(x^{5}-1)^{-1}(x^{4}-1)^{-2}(x^{3}-1)^{-1}(x^{2}-1)\nonumber\\
\hspace{-1.in} P^{F+}_{14} &=& (x^{34}-1)^{11}(x^{28}-1)^{11}(x^{23}-1)(x^{20}-1)^{5}(x^{17}-1)^{-11}(x^{14}-1)^{-11}
(x^{11}-1)^{3} \nonumber\\
\hspace{-1.in} &&\qquad (x^{4}-1)^{3}(x^{2}-1)^{-3}(x-1)\nonumber\\
\hspace{-1.in} P^{F+}_{15} &=& (x^{50}-1)(x^{32}-1)^{18}(x^{26}-1)^{8}(x^{22}-1)^{6}(x^{25}-1)^{-1}(x^{19}-1)^{14}
(x^{16}-1)^{-18} \nonumber\\
\hspace{-1.in} &&\qquad (x^{13}-1)^{-8}(x^{10}-1)(x^{8}-1)(x^{4}-1)^{-1}(x^{2}-1)^{-2}(x-1)\nonumber\\
\hspace{-1.in} P^{F+}_{16} &=& (x^{54}-1)(x^{48}-1)^{6}(x^{30}-1)^{17}(x^{27}-1)^{-1}(x^{24}-1)^{-2}(x^{21}-1)^{17}(x^{18}-1)^{26} \nonumber\\
\hspace{-1.in} &&\qquad (x^{15}-1)^{-17}(x^{12}-1)^{-4}(x^{10}-1)^{-1}(x^{7}-1)(x^{6}-1)(x^{5}-1)(x^{3}-1)^{-1} \nonumber\\
\hspace{-1.in} &&\qquad (x^{2}-1)^{2}(x-1)\nonumber\\
\hspace{-1.in} P^{F+}_{17} &=& (x^{52}-1)^{7}(x^{46}-1)^{21}(x^{29}-1)(x^{28}-1)^{11}(x^{26}-1)^{-7}(x^{23}-1)^{-21}
(x^{22}-1) \nonumber\\
\hspace{-1.in} &&\qquad (x^{20}-1)^{38}(x^{17}-1)^{31}(x^{14}-1)^{-11}(x^{11}-1)^{-1}(x^{10}-1)^{-1}(x^{8}-1) \nonumber\\
\hspace{-1.in} &&\qquad (x^{4}-1)^{-3}(x^{2}-1)(x-1)\nonumber\\
\hspace{-1.in} P^{F+}_{18} &=& (x^{50}-1)^{25}(x^{44}-1)^{50}(x^{31}-1)(x^{28}-1)^{7}(x^{26}-1)^{4}(x^{25}-1)^{-25}
(x^{22}-1)^{-50} \nonumber\\
\hspace{-1.in} &&\qquad (x^{19}-1)^{52}(x^{16}-1)^{26}(x^{13}-1)^{-4}(x^{8}-1)^{-1}(x^{4}-1)^{4}(x^{2}-1)^{-3}\nonumber\\
\hspace{-1.in} P^{F+}_{19} &=& (x^{66}-1)(x^{48}-1)^{67}(x^{42}-1)^{82}(x^{33}-1)^{-1}(x^{30}-1)^{8}(x^{27}-1)^{29}
(x^{24}-1)^{-66} \nonumber\\
\hspace{-1.in} &&\qquad (x^{21}-1)^{-82}(x^{18}-1)^{52}(x^{15}-1)^{13}(x^{14}-1)^{-1}(x^{12}-1)^{-1}(x^{9}-1) \nonumber\\
\hspace{-1.in} &&\qquad (x^{8}-1)(x^{7}-1)(x^{6}-1)^{2}(x^{5}-1)(x^{4}-1)^{-1}(x^{3}-1)^{-1}(x^{2}-1)^{-2}\nonumber\\
\hspace{-1.in} P^{F+}_{20} &=& (x^{70}-1)(x^{64}-1)^{8}(x^{46}-1)^{122}(x^{40}-1)^{97}(x^{35}-1)^{-1}(x^{32}-1)^{-8}(x^{29}-1)^{34} \nonumber\\
\hspace{-1.in} &&\qquad (x^{26}-1)^{85}(x^{23}-1)^{-122}(x^{20}-1)^{-97}(x^{17}-1)^{35}(x^{14}-1)^{5}(x^{8}-1)^{-1} \nonumber\\
\hspace{-1.in} &&\qquad (x^{4}-1)(x^{2}-1)^{3}
\end{eqnarray}
and from these we find that the degrees of the multiplicity of the
eigenvalue $+1$ are 
\begin{equation}
\hspace{-1.in} 1, 0, 0, 1, 1, 2, 1, 2, 1, 4, 7, 11, 8, 10, 20, 47, 69, 86, 103, 162, \ldots
\end{equation}
where again there is a ``mod 4'' effect.

\subsection{Characteristic polynomials $P^C_{L_h}$}

The characteristic polynomials $P^C_{L_h}$ for $T_C(-1;L_h)$ have been well
analyzed in \cite{jon1} and are listed in appendix A of that paper to
$L_h=50$. The degree of the polynomials are the Lucas numbers
which satisfy the recursion relation (\ref{fib}) with initial
conditions $L(0)=2,~L(1)=1$
and which have the generating function
\begin{equation}
G^C = \frac{1+2t}{(1-t-t^2)}
\end{equation}
From the long list of \cite{jon1} we
find that the degrees of the  multiplicity of the eigenvalue $+ 1$ are
\begin{eqnarray}
\hspace{-1.in}  1,1,2,3,0,4,1,7,8,13,2,26,9,49,38,107,28,228,49,501,324,1101,258,2766,469, \nonumber\\
\hspace{-1.in} 5845,3790,13555,2376,35624,5813,75807,38036,180213,30482,480782,69593, \nonumber\\\hspace{-1.in} 1047429,485658,2542453,385020,6794812,914105,15114481,9570844,36794329,\nonumber\\
\hspace{-1.in} 5212354,101089306,12602653,222317557, \ldots
\end{eqnarray}
where we find a ``mod 6'' effect.

\subsection{Characteristic polynomials $P_{L_h}^{C0+}$}

The degrees of the polynomials $P_{L_h}^{C0+}$ are discussed in appendix B of
\cite{ree} and are the series A129526  in the
OEIS~\cite{oeis}. However, an explicit form is not known.

We have computed the
characteristic polynomials in the less
restrictive case of the momentum $P=0$ sector.  The degrees of the 
polynomials follow the
series A000358 in the OEIS~\cite{oeis}, 
which is given by the formula
\begin{equation}
\frac{1}{L_h}\sum_{n|L_h}\phi\left(\frac{L_h}{n}\right)
[F(n-2)+F(n)]
\label{p0}
\end{equation}
where $\phi(n)$ is Euler's totient function (the number of
positive integers $<n$ which are relatively prime with $n$).
In particular when $L_h$ is prime (\ref{p0}) specializes to
\begin{equation}
1+\frac{F(L_h-2)+F(L_h)-1}{L_h}
\end{equation}
which grows as $N_G^{L_h}$.

The order of the restricted positive parity  polynomial $P^{C0+}$ is
greater than the negative parity polynomial $P^{C0-}$ and thus
$P^{C0+}$ also grows as  $N_G^{L_h}$.

The first 29 polynomials are
\begin{eqnarray}
\hspace{-1.in} P^{C0+}_{1} &=& (x-1)\nonumber\\
\hspace{-1.in} P^{C0+}_{2} &=& (x^{4}-1)(x^{2}-1)^{-1}\nonumber\\
\hspace{-1.in} P^{C0+}_{3} &=& (x^{3}-1)(x-1)^{-1}\nonumber\\
\hspace{-1.in} P^{C0+}_{4} &=& (x^{2}-1)^{2}(x-1)^{-1}\nonumber\\
\hspace{-1.in} P^{C0+}_{5} &=& (x^{2}-1)^{2}(x-1)^{-1}\nonumber\\
\hspace{-1.in} P^{C0+}_{6} &=& (x^{4}-1)(x^{3}-1)(x^{2}-1)^{-1}\nonumber\\
\hspace{-1.in} P^{C0+}_{7} &=& (x^{4}-1)^{2}(x^{2}-1)^{-2}(x-1)\nonumber\\
\hspace{-1.in} P^{C0+}_{8} &=& (x^{10}-1)(x^{5}-1)^{-1}(x-1)(x^{2}-1)\nonumber\\
\hspace{-1.in} P^{C0+}_{9} &=& (x^{3}-1)(x-1)^{2}(x^{2}-1)^{2}\nonumber\\
\hspace{-1.in} P^{C0+}_{10} &=& (x^{8}-1)(x^{7}-1)(x^{2}-1)^{-1}(x-1)\nonumber\\
\hspace{-1.in} P^{C0+}_{11} &=& (x^{5}-1)^{2}(x^{4}-1)^{3}(x^{2}-1)^{-3}\nonumber\\
\hspace{-1.in} P^{C0+}_{12} &=& (x^{18}-1)(x^{9}-1)^{-1}(x^{6}-1)(x^{4}-1)(x^{3}-1)^{2}(x^{2}-1)(x-1)^{-1}\nonumber\\
\hspace{-1.in} P^{C0+}_{13} &=& (x^{7}-1)^{3}(x^{2}-1)^{6}(x-1)^{-2}\nonumber\\
\hspace{-1.in} P^{C0+}_{14} &=& (x^{16}-1)^{3}(x^{11}-1)(x^{8}-1)^{-3}(x^{5}-1)(x^{4}-1)(x^{2}-1)^{3}(x-1)^{-1}\nonumber\\
\hspace{-1.in} P^{C0+}_{15} &=& (x^{9}-1)^{4}(x^{6}-1)^{2}(x^{4}-1)^{4}(x^{3}-1)^{3}(x^{2}-1)^{-4}(x-1)^{-1}\nonumber\\
\hspace{-1.in} P^{C0+}_{16} &=& (x^{26}-1)(x^{14}-1)^{3}(x^{13}-1)^{-1}(x^{10}-1)^{3}(x^{7}-1)^{-2}(x^{5}-1)^{3}(x^{4}-1)^{5} \nonumber\\
\hspace{-1.in} &&\qquad (x^{2}-1)^{-4}(x-1)\nonumber\\
\hspace{-1.in} P^{C0+}_{17} &=& (x^{11}-1)^{6}(x^{8}-1)^{8}(x^{4}-1)^{-2}(x^{2}-1)^{4}(x-1)^{3}\nonumber\\
\hspace{-1.in} P^{C0+}_{18} &=& (x^{24}-1)^{6}(x^{18}-1)^{3}(x^{15}-1)(x^{12}-1)^{-5}(x^{9}-1)^{3}(x^{6}-1)^{-1}(x^{4}-1)  \nonumber\\
\hspace{-1.in} &&\qquad 
(x^{3}-1)^{4}(x^{2}-1)^{8}(x-1)^{3}\nonumber\\
\hspace{-1.in} P^{C0+}_{19} &=& (x^{13}-1)^{8}(x^{10}-1)^{18}(x^{7}-1)^{3}(x^{5}-1)^{-6}(x^{4}-1)^{5}(x^{2}-1)^{-3}(x-1)^{2}\nonumber\\
\hspace{-1.in} P^{C0+}_{20} &=& (x^{34}-1)(x^{22}-1)^{15}(x^{17}-1)^{-1}(x^{16}-1)^{2}(x^{14}-1)^{6}(x^{11}-1)^{-8}(x^{8}-1) \nonumber\\
\hspace{-1.in} &&\qquad 
(x^{7}-1)^{4}(x^{4}-1)^{17}(x^{2}-1)^{-12}\nonumber\\
\hspace{-1.in} P^{C0+}_{21} &=& (x^{15}-1)^{10}(x^{12}-1)^{27}(x^{9}-1)^{12}(x^{6}-1)^{3}(x^{5}-1)(x^{4}-1)^{9}(x^{3}-1)^{7} \nonumber\\
\hspace{-1.in} &&\qquad (x^{2}-1)^{-1}(x-1)^{-3}\nonumber\\
\hspace{-1.in} P^{C0+}_{22} &=& (x^{32}-1)^{10}(x^{26}-1)^{14}(x^{20}-1)^{15}(x^{19}-1)(x^{16}-1)^{-10}(x^{13}-1)^{14} \nonumber\\
\hspace{-1.in} &&\qquad 
(x^{10}-1)^{-9}(x^{7}-1)(x^{5}-1)^{6}(x^{4}-1)^{2}(x^{2}-1)^{22}(x-1)^{-2}\nonumber\\
\hspace{-1.in} P^{C0+}_{23} &=& (x^{17}-1)^{13}(x^{14}-1)^{45}(x^{11}-1)^{43}(x^{8}-1)^{4}(x^{7}-1)^{15}(x^{4}-1)^{5} \nonumber\\
\hspace{-1.in} &&\qquad (x^{2}-1)^{16}(x-1)^{-3}\nonumber\\
\hspace{-1.in} P^{C0+}_{24} &=& (x^{42}-1)(x^{30}-1)^{45}(x^{24}-1)^{27}(x^{21}-1)^{-1}(x^{18}-1)^{16}(x^{15}-1)^{-20} \nonumber\\
\hspace{-1.in} &&\qquad (x^{12}
-1)^{9}(x^{10}-1)^{10}(x^{9}-1)^{2}(x^{8}-1)^{3}(x^{6}-1)^{9}(x^{5}-1)^{-10} \nonumber\\
\hspace{-1.in} &&\qquad (x^{4}-1)^{27}(x^{3}-1)^{12}(x^{2}-1)^{-23}\nonumber\\
\hspace{-1.in} P^{C0+}_{25} &=& (x^{19}-1)^{16}(x^{16}-1)^{92}(x^{13}-1)^{116}(x^{10}-1)^{20}(x^{8}-1)^{-8}(x^{5}-1)^{5} \nonumber\\
\hspace{-1.in} &&\qquad (x^{4}-1)^{41}(x^{2}-1)^{-33}(x-1)^{2}\nonumber\\
\hspace{-1.in} P^{C0+}_{26} &=& (x^{40}-1)^{15}(x^{34}-1)^{42}(x^{28}-1)^{105}(x^{23}-1)(x^{22}-1)^{20}(x^{20}-1)^{-15} \nonumber\\
\hspace{-1.in} &&\qquad (x^{17}-1)^{36}(x^{16}-1)(x^{14}-1)^{-45}(x^{11}-1)^{9}(x^{8}-1)^{-1} \nonumber\\
\hspace{-1.in} &&\qquad (x^{7}-1)^{28}(x^{4}-1)^{16}(x^{2}-1)^{26}(x-1)^{4}\nonumber\\
\hspace{-1.in} P^{C0+}_{27} &=& (x^{21}-1)^{19}(x^{18}-1)^{155}(x^{15}-1)^{263}(x^{12}-1)^{92}(x^{9}-1)^{-27}(x^{7}-1) \nonumber\\
\hspace{-1.in} &&\qquad (x^{5}-1)^{26}(x^{4}-1)^{17}(x^{3}-1)^{19}(x^{6}-1)^{7}(x-1)^{9}(x^{2}-1)^{67}\nonumber\\
\hspace{-1.in} P^{C0+}_{28} &=& (x^{50}-1)(x^{38}-1)^{120}(x^{32}-1)^{168}(x^{26}-1)^{110}(x^{25}-1)^{-1}(x^{22}-1)^{15} \nonumber\\
\hspace{-1.in} &&\qquad (x^{20}-1)^{5}(x^{19}-1)^{-54}(x^{16}-1)^{42}(x^{13}-1)^{-26}(x^{11}-1)^{6} \nonumber\\
\hspace{-1.in} &&\qquad (x^{10}-1)(x^{8}-1)^{43}(x
^{5}-1)^{4}(x^{4}-1)^{55}(x^{2}-1)^{-10}(x-1)^{2}\nonumber\\
\hspace{-1.in} P^{C0+}_{29} &=& (x^{23}-1)^{23}(x^{20}-1)^{205}(x^{17}-1)^{581}(x^{14}-1)^{364}(x^{11}-1)^{36}
(x^{7}-1)^{-14} \nonumber\\
\hspace{-1.in} &&\qquad (x^{10}-1)^{15}(x^{5}-1)^{-5}(x^{4}-1)^{131}(x^{2}-1)^{-115}
\end{eqnarray}
and from these we find that the degrees of the multiplicity of the
eigenvalue $+1$ are
\begin{eqnarray}
\hspace{-1.in} 1,2,2,3,3,5,5,8,9,14,16,26,31,49,64,99,133,209,291,455,657,
      1022,1510,2359,\nonumber\\
\hspace{-1.in}  3545,5536,8442,13201,20319,31836,49353,77436,
      120711,189674,296854,467160,\nonumber\\
      \hspace{-1.in} 733363,1155647,1818594,2869378,4524081,7146483, \ldots
\end{eqnarray}
where we find there is a ``mod 6'' effect.

\section{Partition functions at $z=-1$} 
\label{appb}

Successive powers of transfer matrices always satisfy a linear recursion relation, since any matrix satisfies its own characteristic polynomial. Therefore, any linear function of the matrix or its components which is independent of the power of the matrix will also satisfy the same linear recursion relation. The usual functions involved in creating partition functions from transfer matrices, the trace of the matrix, dot products with boundary vectors, and modified traces to account for M{\"o}bius and Klein bottle boundary conditions, all cause the respective partition functions to satisfy the same linear recursion relation as their transfer matrix, its characteristic polynomial. In particular, the Klein bottle partition function $Z^{KC}_{L_v,L_h}(z)$ satisfies the same linear recursion relation in $L_v$ as the torus $Z^{CC}_{L_v,L_h}(z)$, since it is constructed from the same transfer matrix $T_C(z;L_h)$, and the cylinder partition function $Z^{CF}_{L_v,L_h}(z)$ satisfies the same recursion relation in $L_v$ as the M{\"o}bius partition function $Z^{MF}_{L_v,L_h}(z)$ since they are both constructed from the same transfer matrix $T_F(z;L_h)$.

Therefore, the generating functions for the partition functions for a given $L_h$ and for general $z$ are rational functions in $z$ and $x=L_v$ whose denominators are the characteristic polynomials of the $L_h$ transfer matrix and whose numerators are polynomials given by the product of the characteristic polynomial and the initial terms of the series (the numerator has degree 1 less in $x$ than the degree of the characteristic polynomial). 

When the transfer matrix can be block diagonalized and the boundary vector dot products cause the partition function to be a function of only a restricted set of matrices in the direct sum, the partition function will satisfy a recursion relation of smaller order than the order of the full transfer matrix. As an example, $T_C(z;L_h)$ can be block diagonalized into different momentum sectors, and $Z^{FC}_{L_v,L_h}(z)$ is only a function of the reflection symmetric zero momentum sector $0^+$, so that the cylinder $Z^{FC}_{L_v,L_h}(z)$ will satisfy a recursion relation in $L_v$ of the order of the $0^+$ sector and not the order of the full $T_C(z;L_h)$ matrix. Likewise, $Z^{FF}_{L_v,L_h}(z)$ satisfies a recursion relation in $L_v$ of the order of the positive parity sector of $T_F(z;L_h)$. 

Beyond restrictions to particular matrix sectors, however, in general the polynomials in the numerator and denominator of the generating functions do not partially cancel, regardless of the initial conditions of the recursion relation, so that partition functions in $z$ generally satisfy a recursion relation of the same order as its transfer matrix. This holds generically for hard hexagons and hard squares even if at particular values of $z$ some cancellations can occur in the generating function.

For hard squares at $z=-1$ the denominators of the generating
functions simplify to the expressions given in appendix A, whose
orders grow according to the order of the transfer matrices. The
numerators, however, are such that massive cancellations occur, 
so that the partition functions as a function of $x=L_v$ at $z=-1$
satisfy linear recursion relations of much smaller degree than than
the partition function does for general $z$. The form of the numerator
is dependent on the initial conditions of the recursion relation, that
is, the partition function value at $z=-1$ for the first several
values of $L_v$. This, in turn, is dependent on boundary conditions:
both the torus and the Klein bottle partition functions satisfy the
same linear recursion relation of their transfer matrix $T_C(-1;L_h)$,
 but the numerators of their generating functions are different, so that the Klein bottle exhibits much more massive cancellations than the torus for a given $L_h$. Likewise, the cylinder $Z^{CF}_{L_v,L_h}(-1)$ and the M{\"o}bius band have different recursion relation orders due to different amounts of cancellations at $z=-1$. 

The cylinder has the property that for odd $L_h$,
$Z^{FC}_{L_v,L_h}(-1)=-2$ whenever
$\mathrm{gcd}(L_h-1,L_v)=0\quad\mathrm{mod}~3$, and
$Z^{FC}_{L_v,L_h}(-1)=1$ otherwise~\cite{jon3}. Therefore, the linear
recursion relation of $Z^{FC}_{L_v,L_h}(-1)$ for odd $L_h$ is always
of order 1 or 2, even though for generic $z$ the partition function
$Z^{FC}_{L_v,L_h}(z)$ satisfies a linear recursion relation of the
order of the $0^+$ sector of the $T_C(z;L_h)$ transfer matrices, which
grows as $N_G^{L_h}$. The initial conditions for the cylinder for odd $L_h$, therefore, are able to effect incredible cancellations to its generating function whose denominators are given in Appendix A.

In~\cite{jon1} it was proven that for the torus partition function,
$Z^{CC}_{L_v,L_h}(-1)=1$ whenever $L_v,L_h$ are co-prime. Since for
each $L_h$ the torus at $z=-1$ satisfies a linear recursion relation,
its initial conditions happen to be exactly suited to allow for this
number theoretic property. This property does not extend to other
boundary conditions even when they satisfy the same overall linear
recursion relation. The Klein bottle satisfies the same $T_C(-1;L_h)$
linear recursion relation that the torus also satisfies, but its
initial conditions do not cause it to share in the torus' co-primality
property.

A repeating sequence with period $n$ will have a generating function
of the form $p(x)/(1-x^n)$. Therefore, since all of the eigenvalues of
the transfer matrices $T_C(-1
;L_h)$ and $T_F(-1;L_h)$ are roots of unity, as long as the
denominators have only square-free factors, the sequences of partition
function values at $z=-1$ will be repeating, with a period given by
the lcm of the exponents $n_j$ in the factors $(1-x^{n_j})$. Most
sequences below are repeating, with a period often much larger than
the order of the transfer matrix. For the limited cases considered
below, all generating functions along a periodic direction (including
a twist for M{\"o}bius bands and Klein bottles) are repeating. Along
the free direction, the sequences are not always repeating; the
cylinder for $L_h=0$ mod 4 is non-repeating and the free-free
partition function is non-repeating for four of the $L_h$
considered. In~\cite{adam} a general form for the generating functions
of $Z^{FC}_{L_v,L_h}(-1)$ for even $L_h$ is conjectured, 
along with the conjecture that for even $L_h$ the only repeating
sequences for $Z^{FC}_{L_v,L_h}(-1)$ are when $L_h=2$ mod 4. 
We make the following conjecture: 
\newtheorem{conj}{Conjecture}
\begin{conj}
Along a periodic direction (including twists) all generating functions are repeating. 
\end{conj}
We further find below that along the periodic direction, all repeating
sequences are sums of repeating sub-sequences of period $p_j$ which
have value zero except at locations $p_j-1$ mod $p_j$ where their
value is an integer multiple of $p_j$. Often the value is exactly
$p_j$. Therefore, the generating functions along a periodic direction
are logarithmic derivatives of a product of factors of the form $(1-x^{p_j})^{m_j}$, where $m_j$ is an integer. We conjecture that this always holds:
\begin{conj}
Along a periodic direction (including twists), all generating
functions are logarithmic derivatives of products of the form $\prod_j
(1-x^{p_j})^{m_j}$, where $p_j$ and $m_j$ are integers.
\end{conj}
As it turns out, for the limited cases considered below, we find surprisingly that the generating functions for the torus and the cylinder along the periodic direction are exactly the negative of the logarithmic derivative of the characteristic polynomial of their transfer matrices at $z=-1$, so that we have the further conjecture:
\begin{conj}
The generating functions of the torus and cylinder (along the periodic
direction) are equal to the negative of the logarithm of their
characteristic polynomials, that is,
$G^{CC}_{L_h}=-\frac{d}{dx}\ln\left(P^{CC}_{L_h}\right)$ and
$G^{CF}_{L_h}=-\frac{d}{dx}\ln\left(P^{CF}_{L_h}\right)$, respectively.
\end{conj}
This is similar to a conjecture in~\cite{aab}. We note that this
does not hold for general $z$, nor for M{\"o}bius bands or Klein
bottles at $z=-1$. Due to this conjecture, we can use the results from appendix~\ref{appa} to further the tables of periods for the sequences
$Z^{CC}_{L_v,L_h}(-1)$ and $Z^{CF}_{L_v,L_h}(-1)$, where we notice 
a mod 3 pattern. 

For $Z^{CC}_{L_v,L_h}(-1)$, for $L_h=0$~mod~3 we conjecture that the periods are given by the $\mathrm{lcm}(L_h, 2L_h, \ldots, nL_h)$, where $n$ is often given by $n=L_h/3-1$.

For $Z^{CF}_{L_v,L_h}(-1)$, for $L_h=1$~mod~3 we conjecture that the
periods are given by the $\mathrm{lcm}(6,12, \ldots, 6n)$, where $n$
is often $2(L_h-4)/3+1$.

We also note that the periods of the cylinder (along the periodic
$L_v$ direction), the M{\"o}bius band, and the free-free plane are all
equal, and the periods of the Klein bottle and cylinder (along the
free $L_v$ direction) are equal.

Below we list both the generating functions and tables of values for
all boundary conditions, since number theoretic properties such as the
torus's co-primality property can be missed by simply considering the
generating functions. The periods of repeating sequences are
tabulated, along with the minimal order of the recursion relations. 
All generating functions listed were determined by computing all 
partition function values up to the order of the transfer matrix and 
canceling the numerator and denominators of the generating function 
to arrive at the minimal order linear recursion relation; however, 
we extend the table of values to higher $L_h$.

\subsection{The torus $Z^{CC}_{Lv,L_h}(-1)$}
\begin{table}[!ht]
\hspace{-0.5in}
\begin{scriptsize}
\begin{tabular}{|c|rrrrrrrrrrrrrrrrrrrr|}
\hline
$\mathbf{L_h\backslash L_v}$   & \textbf{1} & \textbf{2} & \textbf{3} & \textbf{4} & \textbf{5} & \textbf{6} & \textbf{7} & \textbf{8} & \textbf{9} & \textbf{10} & \textbf{11} & \textbf{12} & \textbf{13} & \textbf{14} & \textbf{15} & \textbf{16} & \textbf{17} & \textbf{18} & \textbf{19} & \textbf{20} \\
\hline
\textbf{1} & 1 & 1 & 1 & 1 & 1 & 1 & 1 & 1 & 1 & 1 & 1 & 1 & 1 & 1 & 1 & 1 & 1 & 1 & 1 & 1  \\
\textbf{2}  & 1 & -1 & 1 & 3 & 1 & -1 & 1 & 3 & 1 & -1 & 1 & 3 & 1 & -1 & 1 & 3 & 1 & -1 & 1 & 3  \\
\textbf{3}  &  1 & 1 & 4 & 1 & 1 & 4 & 1 & 1 & 4 & 1 & 1 & 4 & 1 & 1 & 4 & 1 & 1 & 4 & 1 & 1  \\
\textbf{4}  & 1 & 3 & 1 & 7 & 1 & 3 & 1 & 7 & 1 & 3 & 1 & 7 & 1 & 3 & 1 & 7 & 1 & 3 & 1 & 7 \\
\textbf{5}  & 1 & 1 & 1 & 1 & -9 & 1 & 1 & 1 & 1 & 11 & 1 & 1 & 1 & 1 & -9 & 1 & 1 & 1 & 1 & 11 \\
\textbf{6}  & 1 & -1 & 4 & 3 & 1 & 14 & 1 & 3 & 4 & -1 & 1 & 18 & 1 & -1 & 4 & 3 & 1 & 14 & 1 & 3 \\
\textbf{7}  & 1 & 1 & 1 & 1 & 1 & 1 & 1 & 1 & 1 & 1 & 1 & 1 & 1 & -27 & 1 & 1 & 1 & 1 & 1 & 1 \\
\textbf{8}  & 1 & 3 & 1 & 7 & 1 & 3 & 1 & 7 & 1 & 43 & 1 & 7 & 1 & 3 & 1 & 7 & 1 & 3 & 1 & 47 \\
\textbf{9}  & 1 & 1 & 4 & 1 & 1 & 4 & 1 & 1 & 40 & 1 & 1 & 4 & 1 & 1 & 4 & 1 & 1 & 76 & 1 & 1 \\
\textbf{10}  & 1 & -1 & 1 & 3 & 11 & -1 & 1 & 43 & 1 & 9 & 1 & 3 & 1 & 69 & 11 & 43 & 1 & -1 & 1 & 13 \\
\textbf{11} & 1 & 1 & 1 & 1 & 1 & 1 & 1 & 1 & 1 & 1 & 1 & 1 & 1 & 1 & 1 & 1 & 1 & 1 & 1 & 1 \\
\textbf{12} & 1 & 3 & 4 & 7 & 1 & 18 & 1 & 7 & 4 & 3 & 1 & 166 & 1 & 3 & 4 & 7 & 1 & 126 & 1 & 7 \\
\textbf{13} & 1 & 1 & 1 & 1 & 1 & 1 & 1 & 1 & 1 & 1 & 1 & 1 & -51 & 1 & 1 & 1 & 1 & 1 & 1 & 1 \\
\textbf{14} & 1 & -1 & 1 & 3 & 1 & -1 & -27 & 3 & 1 & 69 & 1 & 3 & 1 & 55 & 1 & 451 & 1 & -1 & 1 & 73 \\
\textbf{15} & 1 & 1 & 4 & 1 & -9 & 4 & 1 & 1 & 4 & 11 & 1 & 4 & 1 & 1 & 174 & 1 & 1 & 4 & 1 & 11 \\
\hline
\end{tabular}
\end{scriptsize}
\caption{$Z^{CC}_{L_v,L_h}(-1)$}
\end{table}
The generating functions $G^{CC}_{L_h}$ as a function of $x=L_v$ are given below.
\begin{eqnarray}
\hspace{-1.in} G^{CC}_1 = \frac{1}{(1-x)}, \qquad G^{CC}_2 = G^{CC}_1 + \frac{4x^3}{(1-x^4)} - \frac{2x}{(1-x^2)}, \qquad G^{CC}_3 = G^{CC}_1 + \frac{3x^2}{(1-x^3)}, && \nonumber \\
\hspace{-1.in} G^{CC}_4 = G^{CC}_2 + \frac{4x}{(1-x^2)}, \qquad G^{CC}_5 = G^{CC}_1 + \frac{20x^9}{(1-x^{10})} - \frac{10x^4}{(1-x^5)}, \qquad    &&\nonumber \\
\hspace{-1.in} G^{CC}_6 = G^{CC}_3  -G^{CC}_1 + G^{CC}_2 + \frac{12x^5}{(1-x^6)}, \qquad G^{CC}_7 = G^{CC}_1 + \frac{56x^{27}}{(1-x^{28})} - \frac{28x^{13}}{(1-x^{14})}, && \nonumber \\
\hspace{-1.in} G^{CC}_8 = G^{CC}_4 + \frac{40x^9}{(1-x^{10})}, \qquad G^{CC}_9 = G^{CC}_3 + \frac{36x^{17}}{(1-x^{18})} + \frac{36x^8}{(1-x^9)}, &&\nonumber\\
\hspace{-1.in} G^{CC}_{10} = G^{CC}_2 + \frac{70x^{13}}{(1-x^{14})} +
\frac{10x^4}{(1-x^5)} + \frac{40x^7}{(1-x^8)}, \nonumber\\ 
\hspace{-1.in} G^{CC}_{11} = G^{CC}_1 + \frac{110x^{54}}{(1-x^{55})} + \frac{176x^{43}}{(1-x^{44})} - \frac{88x^{21}}{(1-x^{22})}.  &&
\end{eqnarray}

\begin{table}[!ht]
\centering
\begin{tabular}{c|ccccccccccc}
$Z^{CC}$ $L_h$ & 1 & 2 & 3 & 4 & 5 & 6 & 7 & 8 & 9 & 10 & 11 \\
\hline
$T_C$ order & 2 & 3 & 4 & 7 & 11 & 18 & 29 & 47 & 76 & 123 & 199 \\
min rec order & 1 & 3 & 3 & 4 & 6 & 8 & 15 & 12 & 18 & 24 & 77 \\
 period & 1 & 4 & 3 & 4 & 10 & 12 & 28 & 20 & 18 & 280 & 220
\end{tabular}
\caption{The minimal order of the recursion relation and the period of the repeating sequence of $Z^{CC}_{L_v,L_h}(-1)$ as a function of $L_v$.}
\end{table}

\subsection{The Klein bottle $Z^{KC}_{L_v,L_h}(-1)$ with twist in $L_v$ direction}
\begin{table}[!ht]
\hspace{-0.5in}
\begin{scriptsize}
\begin{tabular}{|c|rrrrrrrrrrrrrrrrrrrr|}
\hline
$\mathbf{L_h\backslash L_v}$   & \textbf{1} & \textbf{2} & \textbf{3} & \textbf{4} & \textbf{5} & \textbf{6} & \textbf{7} & \textbf{8} & \textbf{9} & \textbf{10} & \textbf{11} & \textbf{12} & \textbf{13} & \textbf{14} & \textbf{15} & \textbf{16} & \textbf{17} & \textbf{18} & \textbf{19} & \textbf{20} \\
\hline
\textbf{1} & 1  & 1  & 1 & 1 & 1 & 1 & 1 & 1 & 1 & 1 & 1 & 1 & 1 & 1 & 1 & 1 & 1 & 1 & 1 & 1  \\
\textbf{2}  & -1 & -3 & -1 & 1 & -1 & -3 & -1 & 1 & -1 & -3 & -1 & 1 & -1 & -3 & -1 & 1 & -1 & -3 & -1 & 1  \\
\textbf{3}  &  -1 & -1 & 2 & -1 & -1 & 2 & -1 & -1 & 2 & -1 & -1 & 2 & -1 & -1 & 2 & -1 & -1 & 2 & -1 & -1  \\
\textbf{4}  &  -1 & 5 & -1 & 1 & -1 & 5 & -1 & 1 & -1 & 5 & -1 & 1 & -1 & 5 & -1 & 1 & -1 & 5 & -1 & 1 \\
\textbf{5}  &  -1 & 3 & -1 & 3 & -1 & 3 & -1 & 3 & -1 & 3 & -1 & 3 & -1 & 3 & -1 & 3 & -1 & 3 & -1 & 3 \\
\textbf{6}  &  1 & -5 & 4 & -1 & 1 & -2 & 1 & -1 & 4 & -5 & 1 & 2 & 1 & -5 & 4 & -1 & 1 & -2 & 1 & -1 \\
\textbf{7}  &  1 & -3 & 1 & 5 & 1 & -3 & 1 & 5 & 1 & -3 & 1 & 5 & 1 & -3 & 1 & 5 & 1 & -3 & 1 & 5 \\
\textbf{8}  & 1 & 7 & 1 & 3 & 1 & 7 & 1 & 3 & 1 & 7 & 1 & 3 & 1 & 7 & 1 & 3 & 1 & 7 & 1 & 3  \\
\textbf{9}  & 1 & 5 & 4 & 5 & 1 & 8 & 1 & 5 & 4 & 5 & 1 & 8 & 1 & 5 & 4 & 5 & 1 & 8 & 1 & 5  \\
\textbf{10}  & -1 & -7 & -1 & -3 & -1 & -7 & 13 & 5 & -1 & -7 & -1 & -3 & -1 & -7 & -1 & 5 & -1 & -7 & -1 & -3  \\
\textbf{11} & -1 & -5 & -1 & 3 & 9 & -5 & -1 & 3 & -1 & 5 & -1 & 3 & -1 & -5 & 9 & 3 & -1 & -5 & -1 & 13 \\
\textbf{12} & -1 & 9 & 2 & 5 & -1 & 12 & -1 & 5 & 2 & 9 & -1 & 8 & -1 & 9 & 2 & 5 & -1 & 12 & -1 & 5 \\
\textbf{13} & -1 & 7 & -1 & 7 & -1 & 7 & 13 & 7 & -1 & 7 & -1 & 7 & -1 & 21 & -1 & 7 & -1 & 7 & -1 & 7 \\
\textbf{14} & 1 & -9 & 1 & 3 & 1 & -9 & 1 & -29 & 1 & 1 & 23 & 3 & 1 & -9 & 1 & 3 & 1 & -9 & 1 & 13 \\
\hline
\end{tabular}
\end{scriptsize}
\caption{$Z^{KC}_{L_v,L_h}(-1)$}
\end{table}
The generating functions $G^{KC}_{L_h}$ as a function of $x=L_v$ are given below.
\begin{eqnarray}
\hspace{-1.in} G^{KC}_{1} = \frac{1}{(1-x)}, \qquad G^{KC}_{2} = -G^{KC}_{1} + \frac{4x^3}{(1-x^4)} - \frac{2x}{(1-x^2)}, \qquad G^{KC}_{3} = -G^{KC}_{1} + \frac{3x^2}{(1-x^3)},  && \nonumber \\
\hspace{-1.in} G^{KC}_{4} = -G^{KC}_{2} - 2G^{KC}_{1} + \frac{4x}{(1-x^2)}, \qquad G^{KC}_{5} = G^{KC}_{4} + G^{KC}_{2} + G^{KC}_{1}, && \nonumber\\ \hspace{-1.in} G^{KC}_{6} = -G^{KC}_{4} + G^{KC}_{3} + G^{KC}_{1}, \qquad G^{KC}_{7} = 2G^{KC}_{2} + 3G^{KC}_{1}, \qquad G^{KC}_{8} = G^{KC}_{4} + 2G^{KC}_{1}, &&\nonumber\\
\hspace{-1.in} G^{KC}_{9} = G^{KC}_{5} + G^{KC}_{3} + 3G^{KC}_{1}, \qquad G^{KC}_{10} = -G^{KC}_{8} - \frac{14x^{13}}{(1-x^{14})} + \frac{8x^7}{(1-x^8)} + \frac{14x^6}{(1-x^7)},  && \nonumber \\
\hspace{-1.in} G^{KC}_{11} = G^{KC}_{7} - 2G^{KC}_{1} + \frac{10x^4}{(1-x^5)}.   &&
\end{eqnarray}

\begin{table}[!ht]
\centering
\begin{tabular}{c|ccccccccccc}
$Z^{KC}$ $L_h$ & 1 & 2 & 3 & 4 & 5 & 6 & 7 & 8 & 9 & 10 & 11 \\
\hline
$T_C$ order & 2 & 3 & 4 & 7 & 11 & 18 & 29 & 47 & 76 & 123 & 199 \\
min rec order & 1 & 3 & 2 & 4 & 2 & 5 & 3 & 4 & 4 & 20 & 7 \\
period & 1 & 4 & 3 & 4 & 2 & 12 & 4 & 4 & 6 & 56 & 20
\end{tabular}
\caption{The minimal order of the recursion relation and the period of the repeating sequence of $Z^{KC}_{L_v,L_h}(-1)$ as a function of $L_v$.}
\end{table}

\subsection{The cylinder $Z^{FC}_{Lv,L_h}(-1) = Z^{CF}_{Lh,L_v}(-1)$}
\begin{table}[!ht]
\hspace{-0.5in}
\begin{scriptsize}
\begin{tabular}{|c|rrrrrrrrrrrrrrrrrrrr|}
\hline
$\mathbf{L_h\backslash L_v}$   & \textbf{1} & \textbf{2} & \textbf{3} & \textbf{4} & \textbf{5} & \textbf{6} & \textbf{7} & \textbf{8} & \textbf{9} & \textbf{10} & \textbf{11} & \textbf{12} & \textbf{13} & \textbf{14} & \textbf{15} & \textbf{16} & \textbf{17} & \textbf{18} & \textbf{19} & \textbf{20} \\
\hline
\textbf{1} & 1 & 1 & 1 & 1 & 1 & 1 & 1 & 1 & 1 & 1 & 1 & 1 & 1 & 1 & 1 & 1 & 1 & 1 & 1 & 1  \\
\textbf{2}  & -1 & -1 & 1 & 1 & -1 & -1 & 1 & 1 & -1 & -1 & 1 & 1 & -1 & -1 & 1 & 1 & -1 & -1 & 1 & 1  \\
\textbf{3}  &  -2 & 1 & 1 & -2 & 1 & 1 & -2 & 1 & 1 & -2 & 1 & 1 & -2 & 1 & 1 & -2 & 1 & 1 & -2 & 1  \\
\textbf{4}  & -1 & 3 & -3 & 5 & -5 & 7 & -7 & 9 & -9 & 11 & -11 & 13 & -13 & 15 & -15 & 17 & -17 & 19 & -19 & 21 \\
\textbf{5}  & 1 & 1 & 1 & 1 & 1 & 1 & 1 & 1 & 1 & 1 & 1 & 1 & 1 & 1 & 1 & 1 & 1 & 1 & 1 & 1 \\
\textbf{6}  & 2 & -1 & 1 & 4 & -1 & -1 & 4 & 1 & -1 & 2 & 1 & 1 & 2 & -1 & 1 & 4 & -1 & -1 & 4 & 1 \\
\textbf{7}  & 1 & 1 & 1 & 1 & 1 & 1 & 1 & 1 & 1 & 1 & 1 & 1 & 1 & 1 & 1 & 1 & 1 & 1 & 1 & 1 \\
\textbf{8}  & -1 & 3 & 5 & 5 & 3 & 7 & 1 & 1 & -1 & 3 & -3 & 5 & 3 & 7 & 1 & 9 & -1 & 3 & -3 & 5 \\
\textbf{9}  & -2 & 1 & 1 & -2 & 1 & 1 & -2 & 1 & 1 & -2 & 1 & 1 & -2 & 1 & 1 & -2 & 1 & 1 & -2 & 1 \\
\textbf{10}  & -1 & -1 & 1 & 1 & 9 & -1 & 1 & 1 & -11 & -1 & 1 & 11 & 9 & -1 & 1 & -9 & -11 & -1 & 11 & 11 \\
\textbf{11} & 1 & 1 & 1 & 1 & 1 & 1 & 1 & 1 & 1 & 1 & 1 & 1 & 1 & 1 & 1 & 1 & 1 & 1 & 1 & 1 \\
\textbf{12} & 2 & 3 & -3 & 8 & -5 & 7 & 8 & 9 & -9 & 14 & -11 & 13 & 2 & 15 & -15 & 8 & -17 & 19 & -4 & 21 \\
\textbf{13} & 1 & 1 & 1 & 1 & 1 & 1 & 1 & 1 & 1 & 1 & 1 & 1 & 1 & 1 & 1 & 1 & 1 & 1 & 1 & 1 \\
\textbf{14} & -1 & -1 & 1 & 1 & -1 & 13 & 1 & 1 & 13 & -1 & 15 & 1 & -1 & -15 & 1 & 15 & -15 & -1 & -13 & 15 \\
\textbf{15} & -2 & 1 & 1 & -2 & 1 & 1 & -2 & 1 & 1 & -2 & 1 & 1 & -2 & 1 & 1 & -2 & 1 & 1 & -2 & 1  \\
\textbf{16} & -1 & 3 & 5 & 5 & 3 & 7 & 1 & 33 & -1 & 3 & 13 & 5 & 3 & 7 & -31 & 9 & -1 & 35 & -3 & 5 \\
\hline
\end{tabular}
\end{scriptsize}
\caption{$Z^{FC}_{L_v,L_h}(-1) = Z^{CF}_{L_h,L_v}(-1)$}
\end{table}
The generating functions $G^{FC}_{L_h}$ as a function of $x=L_v$ are given below. For odd $L_h$ there are only two cases:
\begin{eqnarray}
G^{FC}_{3n\pm1} = \frac{1}{(1-x)} \qquad G^{FC}_{3n} = G^{FC}_{3n\pm1} -\frac{3}{(1-x^3)} && \nonumber
\end{eqnarray}
For even $L_h$:
\begin{eqnarray}
\hspace{-1.in} G^{FC}_2 = G^{FC}_1 - \frac{2(1+x)}{(1-x^4)}, \qquad G^{FC}_4 = \frac{x}{(1-x^2)} - \frac{(1-x)^2}{(1-x^2)^2}, \qquad G^{FC}_6  = -G^{FC}_3 + G^{FC}_2 + G^{FC}_1, && \nonumber\\
\hspace{-1.in}  G^{FC}_{8} = \frac{1}{5}G^{FC}_{4} - \frac{4}{5}\frac{(1+x)(1-x^5)^2}{(1-x^{10})(1-x)^2} + \frac{8}{5}\frac{x(1+x)(x^3+3)(1-x^5)}{(1-x^{10})(1-x)}, && \nonumber \\
\hspace{-1.in}  G^{FC}_{10} = G^{FC}_{2} + \frac{10x(1+x)(1+x^2)}{(1-x^8)} - \frac{10x(x^2+x+1)}{(1-x^7)}, &&\nonumber\\
\hspace{-1.in}  G^{FC}_{12} = \frac{7}{9}G^{FC}_{4} - \frac{4(1-x^6)(1-x^9)}{(1-x^{18})} + \frac{2}{3}\frac{(1+x)(1-x^2)(1-x^3)^2}{(1-x^6)^2} + \frac{(2x^5+2x^4+55x^3+55)}{9(1-x^6)}, &&\nonumber\\
\hspace{-1.in}  G^{FC}_{14} = G^{FC}_{2} + \frac{28x^3p^{FC}_{14}}{(1-x^{16})} - \frac{14x^3(x^7+x^6+x^2+x+1)}{(1-x^{11})}  - \frac{14x^3(1+x)}{(1-x^5)}, &&\nonumber\\
\hspace{-1.in}  G^{FC}_{16} = \frac{243G^{FC}_{4}+2108G^{FC}_{1}}{455}
+ \frac{16(1+x)(1-x^{13})p^{FC}_{16;1}}{13(1-x^{26})}  +
\frac{32(1+x)(1-x^7)p^{FC}_{16;2}}{7(1-x^{14})} +
\frac{8(1-x^2)p^{FC}_{16;3}}{5(1-x^{10})}, &&\nonumber\\
\hspace{-1.in} &&
\end{eqnarray}

\begin{eqnarray}
\hspace{-1.in} p^{FC}_{14} &=& x^{12}+x^{11}+x^7+x^6+x^5+x^2+x+1,  \nonumber\\
\hspace{-1.in} p^{FC}_{16;1} &=& -2x^{11}+6x^9-3x^8+4x^7-9x^6+5x^5-5x^4+9x^3-4x^2+3x-6, \nonumber\\
\hspace{-1.in} p^{FC}_{16;2} &=& -x^5+3x^3-x^2+x-3, \nonumber\\
\hspace{-1.in} p^{FC}_{16;3} &=& 7x^7+7x^6-3x^5+x^4-5x^3+12x^2+6x+10.
\end{eqnarray}
\begin{table}[!ht]
\centering
\begin{tabular}{c|cccccccc}
$Z^{FC}$ $L_h$ & 2 & 4 & 6 & 8 & 10 & 12 & 14 & 16  \\
\hline
$T_{C0^+}$ order & 2 & 3 & 5 & 8 & 14 & 26 & 49 & 99  \\
min rec order & 2 & 3 & 5 & 7 & 13 & 15 & 25 & 29  \\
period & 4 & -- & 12 & -- & 56 & -- & 880 & -- 
\end{tabular}
\caption{The minimal order of the recursion relation and the period of the repeating sequence of $Z^{FC}_{L_v,L_h}(-1)$ as a function of $L_v$.}
\end{table}

The generating functions $G^{CF}_{L_h}$ as a function of $x=L_v$ are given below.
\begin{eqnarray}
\hspace{-1.in} G^{CF}_{1} = \frac{6x^5}{(1-x^6)} - \frac{3x^2}{(1-x^3)} - \frac{2x}{(1-x^2)} + \frac{1}{(1-x)}, \qquad G^{CF}_{2} = \frac{4x^3}{(1-x^4)} - \frac{2x}{(1-x^2)} + \frac{1}{(1-x)},  &&\nonumber\\
\hspace{-1.in} G^{CF}_{3} = \frac{8x^7}{(1-x^8)} - \frac{4x^3}{(1-x^4)} + \frac{1}{(1-x)}, \qquad G^{CF}_{4} = G^{CF}_{1} + G^{CF}_{2} + \frac{4x}{(1-x^2)} - \frac{1}{(1-x)}, &&\nonumber\\
\hspace{-1.in} G^{CF}_{5} = G^{CF}_{3} + \frac{10x^9}{(1-x^{10})} - \frac{2x}{(1-x^2)}, \qquad G^{CF}_{6} = 2G^{CF}_{2} +  \frac{14x^{13}}{(1-x^{14})} + \frac{2x}{(1-x^2)} - \frac{1}{(1-x)}, &&\nonumber\\
\hspace{-1.in} G^{CF}_{7} = G^{CF}_{3} - G^{CF}_{2} + G^{CF}_{1} + \frac{18x^{17}}{(1-x^{18})} + \frac{12x^{11}}{(1-x^{12})}, &&\nonumber\\
\hspace{-1.in} G^{CF}_{8} = -G^{CF}_{3} + \frac{22x^{21}}{(1-x^{22})} + \frac{32x^{15}}{(1-x^{16})} + \frac{4x^3}{(1-x^4)} + \frac{2}{(1-x)}, &&\nonumber\\
\hspace{-1.in} G^{CF}_{9} = G^{CF}_{6} - G^{CF}_{5} + G^{CF}_{2} + \frac{26x^{25}}{(1-x^{26})} + \frac{60x^{19}}{(1-x^{20})} + \frac{16x^7}{(1-x^8)} - \frac{24x^3}{(1-x^4)}, &&\nonumber\\
\hspace{-1.in} G^{CF}_{10} = G^{CF}_{4} - G^{CF}_{3} + G^{CF}_{2} + \frac{30x^{29}}{(1-x^{30})} + \frac{72x^{23}}{(1-x^{24})} + \frac{36x^{17}}{(1-x^{18})}. &&
\end{eqnarray}

\begin{table}[!ht]
\centering
\begin{tabular}{c|cccccccccc}
$Z^{CF}$ $L_h$ & 1 & 2 & 3 & 4 & 5 & 6 & 7 & 8 & 9 & 10  \\
\hline
$T_F$ order & 2 & 3 & 5 & 8 & 13 & 21 & 34 & 55 & 89 & 144  \\
min rec order & 2 & 3 & 5 & 6 & 13 & 16 & 26 & 36 & 60 & 60  \\
period & 6 & 4 & 8 & 12 & 40 & 28 & 72 & 176 & 3640 & 360
\end{tabular}
\caption{The minimal order of the recursion relation and the period of the repeating sequence of $Z^{CF}_{L_v,L_h}(-1)$ as a function of $L_v$.}
\end{table}

\subsection{The M{\"o}bius band $Z^{MF}_{L_v,L_h}(-1)$ with twist in the $L_v$ direction}
\begin{table}[!ht]
\hspace{-0.5in}
\begin{scriptsize}
\begin{tabular}{|c|rrrrrrrrrrrrrrrrrrrr|}
\hline
$\mathbf{L_h\backslash L_v}$   & \textbf{1} & \textbf{2} & \textbf{3} & \textbf{4} & \textbf{5} & \textbf{6} & \textbf{7} & \textbf{8} & \textbf{9} & \textbf{10} & \textbf{11} & \textbf{12} & \textbf{13} & \textbf{14} & \textbf{15} & \textbf{16} & \textbf{17} & \textbf{18} & \textbf{19} & \textbf{20} \\
\hline
\textbf{1}  & 1 & -1 & -2 & -1 & 1 & 2 & 1 & -1 & -2 & -1 & 1 & 2 & 1 & -1 & -2 & -1 & 1 & 2 & 1 & -1 \\
\textbf{2}  & -1 & -3 & -1 & 1 & -1 & -3 & -1 & 1 & -1 & -3 & -1 & 1 & -1 & -3 & -1 & 1 & -1 & -3 & -1 & 1 \\
\textbf{3}  & -1 & -1 & -1 & -5 & -1 & -1 & -1 & 3 & -1 & -1 & -1 & -5 & -1 & -1 & -1 & 3 & -1 & -1 & -1 & -5 \\
\textbf{4}  & -1 & 3 & -4 & -1 & -1 & 6 & -1 & -1 & -4 & 3 & -1 & 2 & -1 & 3 & -4 & -1 & -1 & 6 & -1 & -1 \\
\textbf{5}  & -1 & 1 & -1 & -3 & 9 & 1 & -1 & 5 & -1 & 1 & -1 & -3 & -1 & 1 & 9 & 5 & -1 & 1 & -1 & -3 \\
\textbf{6}  & 1 & -5 & 1 & 3 & 1 & -5 & 15 & 3 & 1 & -5 & 1 & 3 & 1 & -5 & 1 & 3 & 1 & -5 & 1 & 3 \\
\textbf{7}  & 1 & -3 & -2 & -3 & 1 & 12 & 1 & 5 & -20 & -3 & 1 & 0 & 1 & -3 & -2 & 5 & 1 & 12 & 1 & -3 \\
\textbf{8}  & 1 & 5 & 1 & -3 & 1 & 5 & 1 & -27 & 1 & 5 & -21 & -3 & 1 & 5 & 1 & 5 & 1 & 5 & 1 & -3 \\
\textbf{9}  & 1 & 3 & 1 & -5 & 1 & 3 & -13 & 3 & 1 & -47 & 1 & -5 & 27 & 3 & 1 & 3 & 1 & 3 & 1 & 5 \\
\textbf{10} & -1 & -7 & -4 & -3 & -1 & -4 & -1 & 5 & -40 & -7 & -1 & 72 & -1 & -7 & 26 & 5 & -1 & -4 & -1 & -3 \\
\textbf{11} & -1 & -5 & -1 & -1 & -1 & -5 & -1 & -9 & -1 & -5 & 87 & -1 & -1 & 93 & -1 & 7 & -35 & -5 & -1 & -1 \\
\textbf{12} &  -1 & 7 & -1 & -5 & -1 & 7 & -1 & 3 & -1 & 57 & -1 & -5 & 155 & 7 & -1 & -125 & -1 & 7 & -39 & 5 \\
\hline
\end{tabular}
\end{scriptsize}
\caption{$Z^{MF}_{L_v,L_h}(-1)$}
\end{table}
The generating functions $G^{MF}_{L_h}$ as a function of $x=L_v$ are given below.
\begin{eqnarray}
\hspace{-1.in} G^{MF}_{1} = \frac{6x^5}{(1-x^6)} - \frac{3x^2}{(1-x^3)} - \frac{2x}{(1-x^2)} + \frac{1}{(1-x)}, \qquad G^{MF}_{2} = \frac{4x^3}{(1-x^4)} - \frac{2x}{(1-x^2)} - \frac{1}{(1-x)}, && \nonumber\\
\hspace{-1.in} G^{MF}_{3} =  \frac{8x^7}{(1-x^8)} - \frac{4x^3}{(1-x^4)} - \frac{1}{(1-x)}, \qquad  G^{MF}_{4} = -G^{MF}_{2} + G^{MF}_{1} + \frac{4x}{(1-x^2)} - \frac{3}{(1-x)}, && \nonumber\\
\hspace{-1.in} G^{MF}_{5} =  G^{MF}_{3} - \frac{10x^9}{(1-x^{10})} + \frac{10x^4}{(1-x^5)} + \frac{2x}{(1-x^2)}, && \nonumber\\
\hspace{-1.in} G^{MF}_{6} = 2G^{MF}_{2}  -\frac{14x^{13}}{(1-x^{14})} + \frac{14x^6}{(1-x^7)} - \frac{2x}{(1-x^2)} + \frac{3}{(1-x)}, && \nonumber\\
\hspace{-1.in} G^{MF}_{7} = G^{MF}_{3} + G^{MF}_{2} + G^{MF}_{1} +\frac{18x^{17}}{(1-x^{18})} - \frac{12x^{11}}{(1-x^{12})} -\frac{18x^8}{(1-x^9)} +\frac{12x^5}{(1-x^6)} +\frac{2}{(1-x)}, && \nonumber\\
\hspace{-1.in} G^{MF}_{8} = -3G^{MF}_{3} -5G^{MF}_{2} + \frac{22x^{21}}{(1-x^{22})} +\frac{32x^{15}}{(1-x^{16})} -\frac{22x^{10}}{(1-x^{11})} -\frac{6x}{(1-x^2)} -\frac{7}{(1-x)}, && \nonumber\\
\hspace{-1.in} G^{MF}_{9} = -G^{MF}_{6} + G^{MF}_{3} + G^{MF}_{2} -\frac{26x^{25}}{(1-x^{26})} +\frac{60x^{19}}{(1-x^{20})} +\frac{26x^{12}}{(1-x^{13})} -\frac{50x^9}{(1-x^{10})} -\frac{2x}{(1-x^2)} +\frac{4}{(1-x)}, && \nonumber\\
\hspace{-1.in} G^{MF}_{10} = 2G^{MF}_{7} -G^{MF}_{3} -G^{MF}_{1}
-\frac{30x^{29}}{(1-x^{30})} -\frac{72x^{23}}{(1-x^{24})}
+\frac{30x^{14}}{(1-x^{15})} +\frac{96x^{11}}{(1-x^{12})}
-\frac{24x^5}{(1-x^6)} -\frac{3}{(1-x)}. && \nonumber\\
\hspace{-1.in} &&
\end{eqnarray}
\begin{table}[!ht]
\centering
\begin{tabular}{c|cccccccccc}
$Z^{MF}$ $L_h$ & 1 & 2 & 3 & 4 & 5 & 6 & 7 & 8 & 9 & 10  \\
\hline
$T_F$ order & 2 & 3 & 5 & 8 & 13 & 21 & 34 & 55 & 89 & 144  \\
min rec order & 2 & 3 & 5 & 5 & 13 & 16 & 23 & 35 & 60 & 59  \\
period & 6 & 4 & 8 & 12 & 40 & 28 & 72 & 176 & 3640 & 360
\end{tabular}
\caption{The minimal order of the recursion relation and the period of the repeating sequence of $Z^{MF}_{L_v,L_h}(-1)$ as a function of $L_v$.}
\end{table}

\subsection{The free-free plane $Z^{FF}_{L_v,L_h}(-1)$}
\begin{table}[!ht]
\hspace{-0.5in}
\begin{scriptsize}
\begin{tabular}{|c|rrrrrrrrrrrrrrrrrrrr|}
\hline
$\mathbf{L_h\backslash L_v}$   & \textbf{1} & \textbf{2} & \textbf{3} & \textbf{4} & \textbf{5} & \textbf{6} & \textbf{7} & \textbf{8} & \textbf{9} & \textbf{10} & \textbf{11} & \textbf{12} & \textbf{13} & \textbf{14} & \textbf{15} & \textbf{16} & \textbf{17} & \textbf{18} & \textbf{19} & \textbf{20} \\
\hline
\textbf{1}  &  0 & -1 & -1 & 0 & 1 & 1 & 0 & -1 & -1 & 0 & 1 & 1 & 0 & -1 & -1 & 0 & 1 & 1 & 0 & -1 \\
\textbf{2}  & -1 & -1 & 1 & 1 & -1 & -1 & 1 & 1 & -1 & -1 & 1 & 1 & -1 & -1 & 1 & 1 & -1 & -1 & 1 & 1 \\
\textbf{3}  & -1 & 1 & -1 & -1 & 1 & -1 & 1 & 1 & -1 & 1 & -1 & -1 & 1 & -1 & 1 & 1 & -1 & 1 & -1 & -1 \\
\textbf{4}  & 0 & 1 & -1 & 2 & -1 & 3 & -2 & 3 & -3 & 4 & -3 & 5 & -4 & 5 & -5 & 6 & -5 & 7 & -6 & 7 \\
\textbf{5}  & 1 & -1 & 1 & -1 & 1 & 1 & -1 & 3 & -1 & 1 & 1 & -3 & 3 & -1 & 1 & 3 & -3 & 3 & -1 & -1 \\
\textbf{6}  & 1 & -1 & -1 & 3 & 1 & -3 & 1 & 5 & -1 & -5 & 3 & 5 & -3 & -3 & 5 & 3 & -5 & -1 & 5 & 1 \\
\textbf{7}  & 0 & 1 & 1 & -2 & -1 & 1 & 2 & 3 & -1 & -2 & -1 & -1 & 4 & 3 & -1 & -2 & -3 & 3 & 4 & 1 \\
\textbf{8}  & -1 & 1 & 1 & 3 & 3 & 5 & 3 & 3 & 3 & 3 & -1 & 1 & -3 & -1 & -3 & 1 & -3 & 1 & 1 & 5 \\
\textbf{9}  & -1 & -1 & -1 & -3 & -1 & -1 & -1 & 3 & 3 & 1 & 5 & 1 & 5 & 5 & 1 & 1 & -3 & -5 & -5 & -5 \\
\textbf{10} & 0 & -1 & 1 & 4 & 1 & -5 & -2 & 3 & 1 & 2 & -1 & 3 & -4 & -7 & 7 & 10 & -1 & -7 & -4 & 5 \\
\textbf{11} & 1 & 1 & -1 & -3 & 1 & 3 & -1 & -1 & 5 & -1 & 1 & 1 & -1 & -3 & -1 & 7 & -1 & -3 & -3 & 3 \\
\textbf{12} & 1 & 1 & -1 & 5 & -3 & 5 & -1 & 1 & 1 & 3 & 1 & 5 & -1 & 17 & -1 & 5 & 1 & -1 & -1 & 3 \\
\textbf{13} & 0 & -1 & 1 & -4 & 3 & -3 & 4 & -3 & 5 & -4 & -1 & -1 & 2 & 1 & -1 & 0 & 5 & -3 & 10 & -9 \\
\hline
\end{tabular}
\end{scriptsize}
\caption{$Z^{FF}_{L_v,L_h}(-1)$}
\end{table}
The generating functions $G^{FF}_{L_h}$  as a function of $x=L_v$ are given below.
\begin{eqnarray}
\hspace{-1.in} G^{FF}_{1} = \frac{2x^4 (1+x)}{(1-x^6)} + \frac{1}{(1-x^3)} -\frac{1}{(1-x)}, \qquad G^{FF}_{2} = \frac{2x^2(1+x)}{(1-x^4)} -\frac{1}{(1-x)}, &&\nonumber\\
\hspace{-1.in} G^{FF}_{3} = \frac{2x^3(x-1)(1+x^2)}{(1-x^8)} + \frac{2x}{(1-x^2)} -\frac{1}{(1-x)}, &&\nonumber\\
\hspace{-1.in} G^{FF}_{4} = \frac{1}{3}G^{FF}_{1} -\frac{(1-x)^2}{3(1-x^2)^2} +\frac{x}{3(1-x^2)} +\frac{1}{3(1-x)},  &&\nonumber\\
\hspace{-1.in} G^{FF}_{5} = G^{FF}_{3} + \frac{10x^3(1-x)+4(1-x)^2(2x^2+x+2)}{5(1-x^5)} + \frac{2}{5(1-x)}, &&\nonumber\\
\hspace{-1.in} G^{FF}_{6} = 2G^{FF}_{2}+\frac{2(1-x)(5x^5+10x^4+x^3-x^2+11x+9)}{7(1-x^7)} +\frac{3}{7(1-x)}, &&\nonumber\\
\hspace{-1.in} G^{FF}_{7} = G^{FF}_{3} + \frac{p^{FF}_7 (x^6+x^3+1)(1-x^2)}{3(1-x^{18})} + \frac{(1-x)(1-x^2)^2(x^2+x+1)^2}{3(1-x^6)^2} + \frac{1}{3(1-x)}, &&\nonumber\\
\hspace{-1.in} G^{FF}_{8} = \frac{-3G^{FF}_{4}+G^{FF}_{1}}{11} +\frac{2(1+x)(1-x^{11})p^{FF}_{8;1}}{11(1-x^{22})} +\frac{2(1-x^8)p^{FF}_{8;2}}{(1-x^{16})} -\frac{15}{11(1-x)}, &&\nonumber \\
\hspace{-1.in} G^{FF}_{9} =  G^{FF}_{3}+\frac{p^{FF}_{9;1}}{(1-x^{20})}  +\frac{p^{FF}_{9;2}}{(1-x^{14})} +\frac{2(1-x)p^{FF}_{9;3}}{13(1-x^{13})} +\frac{2}{13(1-x)}, &&\nonumber \\
\hspace{-1.in} G^{FF}_{10} = \frac{1}{9}G^{FF}_{4} +\frac{4p^{FF}_{10;1}}{3(1-x^{18})} +\frac{2p^{FF}_{10;2}}{5(1-x^{15})} -\frac{p^{FF}_{10;3}}{2(1-x^{12})} -\frac{p^{FF}_{10;4}}{90(1-x^6)^2} +\frac{p^{FF}_{10;5}}{90(1-x^6)} +\frac{4}{15(1-x)}, &&\nonumber\\
\hspace{-1.in} G^{FF}_{11} = G^{FF}_{3} +\frac{p^{FF}_{11;1}}{(1-x^{34})} +\frac{p^{FF}_{11;2}}{(1-x^{16})} -\frac{2(1-x^2)p^{FF}_{11;3}}{7(1-x^{14})} +\frac{(1-x)p^{FF}_{11;4}}{11(1-x^{11})} +\frac{27}{77(1-x)}, &&
\end{eqnarray}
\begin{eqnarray}
\hspace{-1.in} p^{FF}_7 = x^9+x^8+2x^6-7x^5+4x^4-4x^3+7x^2-2x+1, &&\nonumber \\
\hspace{-1.in} p^{FF}_{8;1} = 8x^9+9x^7-2x^6+x^5+5x^4-5x^3-x^2+2x-9, &&\nonumber\\
\hspace{-1.in} p^{FF}_{8;2} = -x^7-x^6-2x^5-x^4+x+2, &&\nonumber \\
\hspace{-1.in} p^{FF}_{9;1} = 2(2x^5+x^4-x-2)(1+x^2)(1-x^{10}), &&\nonumber\\
\hspace{-1.in} p^{FF}_{9;2} = 2(-x^5-x^3+x^2-x+1)(1+x)(1-x^7), &&\nonumber\\
\hspace{-1.in} p^{FF}_{9;3} = 14x^{11}+28x^{10}+16x^9+17x^8-8x^7-7x^6+7x^5+21x^4+35x^3+36x^2+11x+12, &&\nonumber\\
\hspace{-1.in} p^{FF}_{10;1} = x(1-x)(x^6+x^3+1)(1-x^6), &&\nonumber \\
\hspace{-1.in} p^{FF}_{10;2} = (1+x)(1-x^3)(2x^{10}-4x^8+4x^7+4x^6-7x^5+x^4+7x^3-4x^2-2x+4), &&\nonumber \\
\hspace{-1.in} p^{FF}_{10;3} = (1-x^6)(5x^5+5x^4-4x^3-7x^2+7x+4), &&\nonumber \\
\hspace{-1.in} p^{FF}_{10;4} = (1-x)(1-x^2)^2(37x^4+86x^3+111x^2+86x+37), &&\nonumber \\
\hspace{-1.in} p^{FF}_{10;5} = (97x^3+97x^2+48x+49)(1-x^2), &&\nonumber \\
\hspace{-1.in} p^{FF}_{11;1} = 2x^2(1+x)(1-x^{17})(x^{13}-2x^9+x^8+x^6+x^5-x^4-x^3-x+2), &&\nonumber \\
\hspace{-1.in} p^{FF}_{11;2} = (1-x^8)(x^7+x^6+x^5-3x^4-x^3+x^2+3x-1), &&\nonumber \\
\hspace{-1.in} p^{FF}_{11;3} = 8x^{11}+8x^{10}+2x^9+9x^8-4x^7-4x^6-3x^5+11x^4+12x^3+12x^2-x+6, &&\nonumber \\
\hspace{-1.in} p^{FF}_{11;4} = -4x^9-8x^8-56x^7-16x^6+24x^5+20x^4-28x^3-32x^2+8x+48. &&
\end{eqnarray}

\begin{table}[!ht]
\centering
\begin{tabular}{c|ccccccccccc}
$Z^{FF}$ $L_h$ & 1 & 2 & 3 & 4 & 5 & 6 & 7 & 8 & 9 & 10 & 11 \\
\hline
$T_{F+}$ order & 2 & 2 & 4 & 5 & 9 & 12 & 21 & 30 & 51 & 76 & 127 \\
min rec order & 2 & 2 & 4 & 5 & 9 & 9 & 17 & 21 & 31 & 35 & 51 \\
period & 6 & 4 & 8 & -- & 40 & 28 & -- & -- &  3640 & -- &  20944
\end{tabular}
\caption{The minimal order of the recursion relation and the period of the repeating sequence of $Z^{FF}_{L_v,L_h}(-1)$ as a function of $L_v$.}
\end{table}

\section{Hard square equimodular curves as $|z| \to \infty$}
\label{appc}

Consider hard squares for a system of width $L_h = 2 L$ sites. 
The boundary conditions can be free or periodic, but not restricted 
by parity or momentum. We wish to show that the transfer matrices 
$T_C(z;L_h)$ and $T_F(z;L_h)$ both have $2L$ branches of equimodular 
curves going out to $|z|\rightarrow\infty$.

Let $A$ (resp.\ $B$) denote the maximally packed state with $L$ 
particles occupying the even (resp.\ odd) numbered sites. Similarly, 
for $k \ll L$, let $A_k$ denote the classes of states having $L-k$ 
particles of which $O(L)$ have positions overlapping with those of $A$ 
and $O(1)$ overlap with those of $B$. More loosely, the states $A_k$ 
have the same order as $A$, up to small local perturbations.
The states $B_k$ are similarly defined from $B$.

To discuss the $|z| \to \infty$ limit we replace $z$ by $z^{-1}$ and 
consider a perturbation theory for $|z| \ll 1$. After division by an 
overall factor, the Boltzmann weight of state $A$ is 1, and each of 
the states in the class $A_k$ have weight $z^k$.

To order zero (i.e., considering only states $A$ and $B$) the 
transfer matrix is the permutation matrix of size $2$, with
eigenvalues $\lambda_1 = 1$ and $\lambda_2 = -1$.

To order $k \ll L$ it is easy to see that the only non-zero matrix 
elements connect an $A$-type state to a $B$-type state and vice
versa. Physically this means that if we start from a state which has 
predominantly particles on the even sublattice, it will remain so 
forever: we stay in the same ordered phase. Mathematically it is 
not hard to see that this implies that the eigenvalues 
$\lambda_1$ and $\lambda_2$ will continue to just differ by an 
overall sign, order by order in perturbation theory. Other 
eigenvalues are $O(z)$, hence play no role since then cannot be 
equimodular with $\lambda_1$ and $\lambda_2$.

The perturbative result $\lambda_1 + \lambda_2 = 0$ breaks down at 
an order $k$ which is sufficiently high to create a domain wall 
across the strip/cylinder/torus between the two different ordered states. 
This happens precisely for $k=L$. It follows that 
$\lambda_1 + \lambda_2 = O(z^L)$, implying that
\begin{equation}
 \lambda_2 / \lambda_1 = -1 + O(z^L) \,.
\end{equation}
To obtain equimodularity, the left-hand side must be on the unit
circle. For $|z| \ll 1$ this will happen when $z^L$ is perpendicular 
to $-1$, so that $\arg(z^L) = \pm \pi/2$. It follows that there are 
$2L$ equimodular curves going out of $z=0$ with the angles
\begin{equation}
 \arg(z) = \frac{(1+2k) \pi}{2L} \mbox{ with $k=0,1,\ldots,2L-1$}.
\end{equation}

\end{document}